\documentclass[%
 reprint,
 amsmath,amssymb,
 aps,
 prb,
]{revtex4-1}

\usepackage{graphicx}
\usepackage{dcolumn}
\usepackage{bm}
\usepackage{hyperref}

\begin{document}

\preprint{APS/123-QED}

\title{Aperiodic dynamical quantum phase transitions in multi-band Bloch Hamiltonian and its origin}

\author{Kaiyuan Cao}%
 \affiliation{%
 Research Center for Intelligent Supercomputing, Zhejiang Lab, Hangzhou 311100, P. R. China
}%

\author{Hao Guo}
 \email{guohao.ph@seu.edu.cn}
 \affiliation{School of Physics, Southeast University, Jiulonghu Campus, Nanjing 211189, P. R. China}

\author{Guangwen Yang}%
 \email{ygw@tsinghua.edu.cn}
\affiliation{Research Center for Intelligent Supercomputing, Zhejiang Lab, Hangzhou 311100, P. R. China}
\affiliation{Department of Computer Science and Technology, Tsinghua University, Haidian District, Beijing 100084, P. R. China}%

\date{\today}

\begin{abstract}
  We investigate the dynamical quantum phase transition (DQPT) in the multi-band Bloch Hamiltonian of the one-dimensional periodic Kitaev model, focusing on quenches from a Bloch band. By analyzing the dynamical free energy and Pancharatnam geometric phase, we show that the critical times of DQPTs deviate from periodic spacing due to the multi-band effect, contrasting with results from two-band models. We propose a geometric interpretation to explain this non-uniform spacing. Additionally, we clarify the conditions needed for DQPT occurrence in the multi-band Bloch Hamiltonian, highlighting that a DQPT only arises when the quench from the Bloch states collapses the band gap at the critical point. Moreover, we establish that the dynamical topological order parameter, defined by the winding number of the Pancharatnam geometric phase, is not quantized but still exhibits discontinuous jumps at DQPT critical times due to periodic modulation. Additionally, we extend our analysis to mixed-state DQPT and find its absence at non-zero temperatures.
\end{abstract}

\maketitle

\section{Introduction}

Recent experimental studies on ultra-cold atoms trapped in optical lattices \cite{Bloch2008, Lewenstein2012, Belsley2013} have revolutionized the study of non-equilibrium dynamics in isolated quantum systems \cite{Polkovnikov2011}. One of the key areas of focus in this field is the time evolution of a quantum system after a sudden global quench, which can be easily carried out in experiments and studied theoretically \cite{Aditi2018}. In such cases, the Loschmidt echo, which indicates the overlap between the eigenstates of the pre- and post-quench Hamiltonians, plays a crucial role \cite{Zurek2005}. The formal analogy between the Loschmidt amplitude and the canonical partition function of an equilibrium system has led to the introduction of the concept of dynamical quantum phase transition (DQPT), which helps to understand the notions of phase and phase transition far from equilibrium \cite{Heyl2013110, Zvyagin201642, Heyl201881}.

The dynamical quantum phase transition (DQPT) is a phenomenon that describes the early-time critical behavior of the Loschmidt echo, $\mathcal{L}(t) = |\mathcal{G}(t)|^2$, during the nonequilibrium dynamical evolution of a quantum system. Here, the Loschmidt amplitude, $\mathcal{G}(t)$, measures the overlap of the time-evolving state with the initial state and is given by
\begin{equation}\label{LA.definition}
  \mathcal{G}(t) = \langle\psi_{0}|\psi(t)\rangle = \langle\psi_{0}|e^{-iHt}|\psi_{0}\rangle.
\end{equation}
To manifest the DQPT, one can define the dynamical free energy density as the rate function of the Loschmidt echo in the thermodynamic limit, i.e., $\lambda(t)=-\lim_{N\rightarrow+\infty}\frac{1}{N}\ln{[\mathcal{L}(t)]}$, or its time derivative, which exhibits cusp-like singularities at critical times. The DQPT has been extensively studied in many quantum systems, including XY chains \cite{Vajna201489, PhysRevE.93.052133, Cao202231, Porta.10.1}, Kitaev honeycomb models \cite{Schmitt201592}, non-integrable models \cite{Karrasch201387, Andraschko201489, Heyl2014113, Kriel.90.125106, Sharma.92.104306}, systems with long-range interactions \cite{Halimeh201796, Homrighausen201796,PhysRevB.95.174305, PhysRevE.96.062118, Dutta201796, Bojan2018120, Halimeh.2.033111}, quantum Potts models \cite{PhysRevB.95.075143}, non-Hermitian systems \cite{Zhou201898, Mondal.2022.106, Mondal.2023.107}, Bose-Einstein condensates \cite{Mehdi2019100}, inhomogeneous systems \cite{Yang201796,PhysRevA.97.033624,Mendl2019100,Cao2020102,Modak2021103,Kuliashov.107.094304, Mishra.53.375301}, periodically driven systems \cite{Yang2019100,Zamani2020102, Shirai.101.013809, Zhou202133,Jafari2021103,NC.12.5108,PhysRevB.105.094304,Jafari2022105}, systems in mixed states \cite{Bhattacharya.96.180303, Heyl.96.180304, Lang.98.134310, Bandyopadhyay.8, Hou.102.104305, Kyaw.101.012111, Mera.97.094110, Sedlmayr.97.045147, Link.125.143602, Hou.104.023303},  and others \cite{Heyl.115.140602, Vajna.91.155127, Tatjana.1.1.003, Lang.121.130603, Huang.122.250401, Jafari.99.054302, Khatun.123.160603, Lahiri.99.174311, Liu.99.104307, Srivastav.100.144203, Gulacsi.101.205135, Meibohm_2023.023034, Wong.105.174307, Wrzessniewski.105.094514, Hashizume.4.013250}. Additionally, several experiments have directly observed DQPTs, including trapped ions simulations \cite{Vogel201714, Jurcevic2017119, Chen.102.042222, Muniz.580.602}, 53-qubit quantum simulations \cite{Zhang2017551}, nuclear magnetic resonance quantum simulators \cite{Nie2020124}, quantum walks of photons \cite{Wang2019122, Xu20209}, and spinor condensate simulations \cite{Tian2020124}. Notably, there is another definition of the DQPT, which examines the asymptotic late-time behavior of the order parameters \cite{Yuzbashyan200696, Barmettler2009102, Eckstein2009103, Sciolla2010105,
 Dziarmaga201059}. Two types of DQPTs have been discovered that are related in the long-range quantum Ising chain \cite{Bojan2018120}. Remarkably, the DQPT can be perfectly characterized by a dynamical topological order parameter (DTOP) \cite{Budich201693}.

Even so, further work is necessary to clarify certain aspects of the DQPT. In particular, we focus on two points of contention: the periodicity of critical times and the condition necessary for DQPT occurrence. While many works have found that the critical times of DQPTs are periodically spaced in the time plane, there exists a class of models that demonstrate clear evidence that the critical times of DQPTs may not be uniformly spaced, including nonintegrable models \cite{Karrasch201387}, quantum spin chains with long-range interactions \cite{Halimeh201796, Homrighausen201796}, systems with multiple bands \cite{Huang2016117, Mendl2019100, Haldar2020101, Sedlmayr2020}, and the quantum system with quasiperiodic potential \cite{Modak2021103}. Given that periodic-spaced critical times are predominantly found in two-band models, we hypothesize that the uneven spacing of critical times may be attributed to the influence of multiple bands. With respect to the condition necessary for DQPT occurrence, many theoretical works indicate that a DQPT requires the quench protocol to cross the critical points of quantum phase transitions (QPTs). However, several studies present counterexamples, demonstrating that a DQPT can occur without crossing a QPT \cite{Vajna201489, Andraschko201489, Halimeh201796, Homrighausen201796}, or that no DQPT occurs even with the crossing of a QPT \cite{Haldar2020101, Cao202231}. Based on the findings of Huang et al. in the real Hofstadter model \cite{Huang2016117}, which suggests that the appearance of the DQPT is linked to changes in topological numbers, we posit that the occurrence of a DQPT requires the quench protocol to cross the gap collapsing point, as the change in topological numbers in the real Hofstadter model arises from the gap collapsing.

In this paper, we chose the Bloch Hamiltonian, based on the periodic Kitaev model, as a typical example to confirm our inferences. The model satisfies the prerequisites of our study, as it has multiple energy bands with inversion symmetry and has both the energy gap and gap collapsing points, which facilitate our comparison to determine the influence of gap collapsing (see Fig.~\ref{energy.spectra}). By calculating the rate function and Pancharatnam geometric phase (PGP) in the four-band Bloch Hamiltonian, we found that the critical times of DQPTs are not periodically spaced due to the deviation of the critical wave vectors caused by the multi-band effect. We provide a geometric interpretation to explain the non-uniformly spaced critical times. Our findings also show that only the type of Bloch state that collapses the band gap at the critical point induces the occurrence of DQPTs after the quench across the critical point. We determine that the criterion for the occurrence of DQPTs in Ref.~\onlinecite{Huang2016117} can also be used in our model (complex, Hermitian Bloch Hamiltonian), not only the real Bloch Hamiltonian. Furthermore, we also discuss the influence of multiple bands on the DTOP and mixed-state DQPT.

The paper is organized as follows: In Section.~\uppercase\expandafter {\romannumeral2}, we introduce the multi-band Bloch Hamiltonian under the periodic effects, and the scheme of a global quantum quench in the Bloch Hamiltonian. In Sections.~\uppercase\expandafter {\romannumeral3}, \uppercase\expandafter {\romannumeral4}, and \uppercase\expandafter {\romannumeral5} we discuss the behaviors of the DQPTs, DTOP, and mixed-state DQPT after sudden quenches in the Bloch Hamiltonian $H_{k}^{l=4}(h)$. We summarize our results in Section.~\uppercase\expandafter {\romannumeral6}. In Appendix B, we also present the results of the six-band Bloch Hamiltonian. There is no significant difference between it and the four-band Bloch Hamiltonian.

\section{The models}

We consider a one-dimensional lattice of particles subject to periodic modulation with particle-hole symmetry, described by a multi-band Hamiltonian \cite{Altland199755}. Assuming a unit lattice constant, the Bloch Hamiltonian for $l$ Bloch bands is given by
\begin{equation}\label{Bloch.H}
  H_{k}^{l}(h,\alpha),
\end{equation}
whose explicit expression can be found in Appendix~A. Here $h$ $(h>0)$ is an external field, and $\alpha$ represents the strength of the periodic modulation. For example, the Bloch Hamiltonian in the period-two case has $l=4$ Bloch bands, and that in the period-three case has $l=6$ Bloch bands (see Appendix~A).

Fig.~\ref{energy.spectra} depicts the energy spectra of the Bloch Hamiltonian $H_{k}^{l=4}(h)$, as functions of $h$ with a fixed value of $\alpha=0.5$. The energy spectra exhibit particle-hole symmetry, resulting in symmetric energy spectra with respect to $\varepsilon_{k}=0$. At the critical point $h_{c}$, the energy gap between the middle two spectra collapses. Based on the relationship between the Kitaev model and the quantum spin chain \cite{Pfeuty1979245, Kitaev_2001},  the critical points can be determined as follows \t: for the Bloch Hamiltonian $H_{k}^{l=4}(h)$, $h_{c}=\sqrt{\alpha}\approx0.7071$.

\begin{figure}
  \centering
  \includegraphics[width=0.95\linewidth]{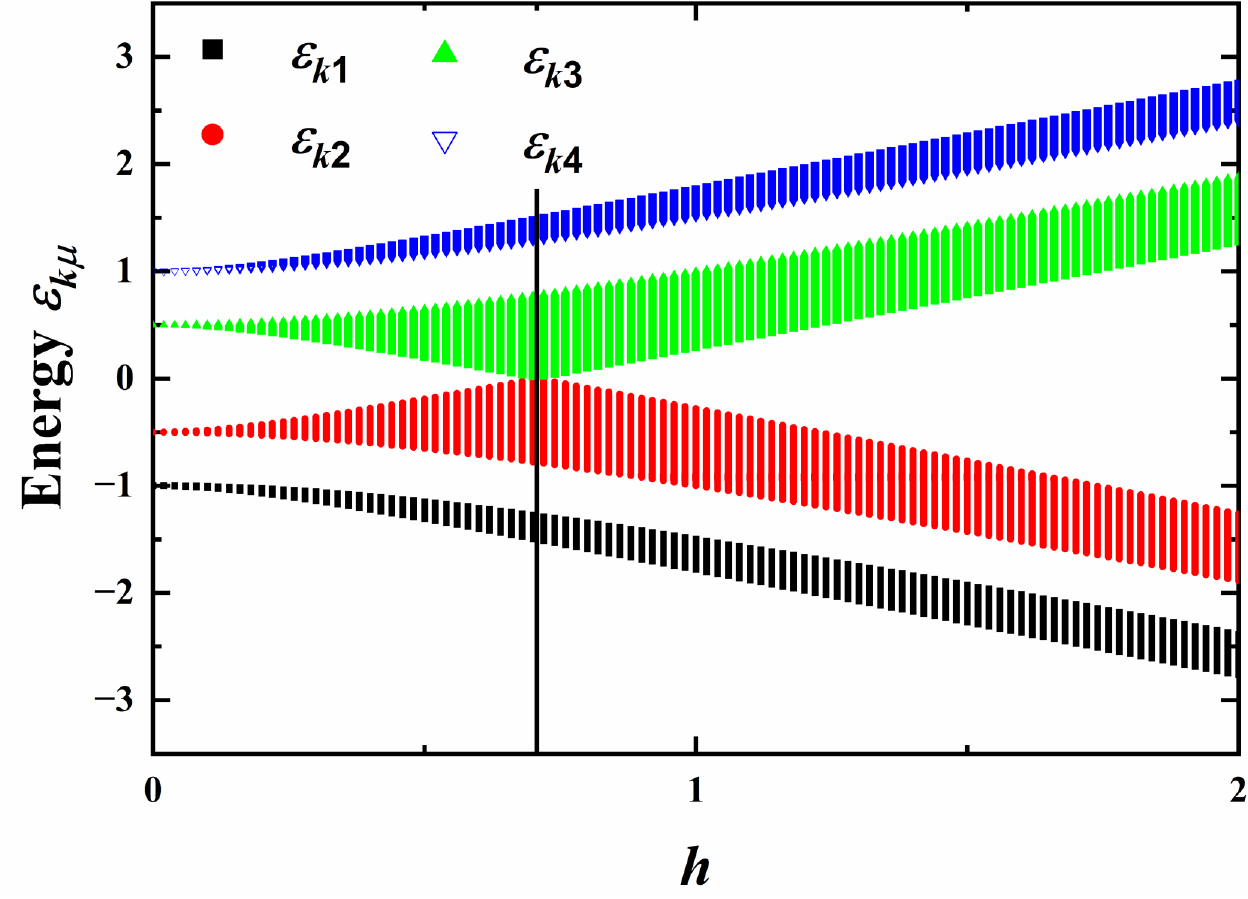}\\
  \caption{ The energy spectra as functions of $h$ for the Bloch Hamiltonian $H_{k}^{l=4}(h)$ with $\alpha=0.5$. In the thermodynamic limit, the energy gaps of the middle two bands close at the critical points $h_{c}=\sqrt{\alpha}\approx0.7071$. 
  }\label{energy.spectra}
\end{figure}

To investigate the dynamics of the Bloch Hamiltonian, we employ a quantum protocol that involves a sudden change of the external field from $h_{0}$ to $h_{1}$. Initially, the system is prepared at a single-filled Bloch state $|u_{k\mu}^{i}\rangle$ $(\mu=1,\cdots,l)$ of the pre-quench Hamiltonian $H_{k}^{l}(h_{0})$, satisfying the eigenvalue problem
\begin{equation}\label{eigen.value.problem}
  H_{k}^{l}(h)|u_{k\nu}\rangle = \varepsilon_{k\nu}|u_{k\nu}\rangle.
\end{equation}
After the quench, the time-evolved state is given by
\begin{equation}\label{time-evolved}
  |\psi_{k}(t)\rangle = e^{-iH_{k}^{l}(h_{1})t}|u_{k\mu}^{i}\rangle = \sum_{\nu=1}^{l}e^{-i\varepsilon_{k\nu}^{f}t}p_{k\nu}|u_{k\nu}^{f}\rangle,
\end{equation}
where the coefficients $p_{k\nu}=\langle u_{k\nu}^{f}|u_{k\mu}^{i}\rangle$ $\nu=1,\cdots,l$ are obtained by expanding the initial state in a linear superposition of the eigenstates of the post-quench Hamiltonian. The probability of adiabatic transition is given by $|p_{k,\nu=\mu}|^{2}$, while the probabilities of nonadiabatic transitions are denoted by$|p_{k,\nu\neq\mu}|^{2}$. The completeness condition requires $\sum_{\nu=1}^{l}|p_{k\nu}|^{2}=1$.

To calculate the Loschmidt amplitude, we substitute Eq.~(\ref{time-evolved}) into Eq.~(\ref{LA.definition}), yielding $\mathcal{G}(t)=\prod_{k>0}\mathcal{G}_{k}(t)$, where $\mathcal{G}_{k}(t)$ is the Loschmidt amplitude for each $k$:
\begin{equation}\label{LA.Bloch}
  \mathcal{G}_{k}(t) = \langle u_{k\mu}^{i}|\psi_{k}(t)\rangle = \sum_{\nu=1}^{l}|p_{k\nu}|^{2}e^{-i\varepsilon_{k\nu}^{f}t}.
\end{equation}
For ease of understanding, we express $\mathcal{G}_{k}(t)$ in terms of polar coordinates as
\begin{equation}\label{LA.polar}
  \mathcal{G}_{k}(t) = r_{k}(t)e^{i\phi_{k}(t)}.
\end{equation}
The Loschmidt echo, also known as the return probability, is thus given by
\begin{equation}\label{Loschmidt.echo}
  \mathcal{L}(t) = \prod_{k>0}\mathcal{L}_{k}(t), \quad\text{where}\quad\mathcal{L}_{k}(t) = |\mathcal{G}_{k}(t)|^{2} = r_{k}^{2}.
\end{equation}
In the thermodynamic limit, DQPTs can be detected by identifying the cusp-like singularities of the dynamical free energy density, which is defined as the rate function of the Loschmidt echo:
\begin{equation}\label{rate.function}
  \lambda(t) = -\lim_{N\rightarrow\infty}\frac{1}{N}\ln{[\mathcal{L}(t)]} = -\int_{0}^{\pi}\frac{dk}{2\pi}\ln{r_{k}^{2}(t)}.
\end{equation}
Here $N$ is the degree of freedom of the system. According to Eqs.~(\ref{LA.Bloch}) and (\ref{Loschmidt.echo}), it is improbable that the Loschmidt amplitude or Loschmidt echo has nonanalytic behaviors with respect to $t$. Therefore, the singularities of the rate function only come from $r_{k}(t)=0$ in the logarithm when DQPTs occur. This indicates that the overlap between the time-evolved states and the initial state vanishes at critical times. In other words, they are orthogonal to each other since they share no common components.

Another straightforward method to depict the DQPT is via the Fisher zeros of the Loschmidt amplitude $\mathcal{G}(z)$ by taking the complex continuation of $\mathcal{G}(t)$, where $z$ is the complexification of $t$ \cite{Heyl2013110, Heyl201881}. This provides an effective way to determine the critical wave vectors $k_{c}$ of DQPTs as well as the distribution of critical times. For instance, in the case of the homogeneous case $H_{k}^{l=2}(h)$, the Fisher zeros are given by \cite{sharma2015,Sharma201693,Zhang_2016114,Divakaran201693}
\begin{equation}\label{Fisher.homogeneous}
  z_{n}(k) = \frac{1}{2\varepsilon_{k}^{f}}\left[\ln{\left(\frac{|p_{k1}|^{2}}{1-|p_{k1}|^{2}}\right)}+i\pi(2n+1)\right],
\end{equation}
which coalesces to a family of lines that intersect with the imaginary axis at critical times $t^*_n$:
\begin{equation}\label{critical.time.homogeneous}
  t_{n}^{*} = (2n+1)t_{0}^{*}, \quad t_{0}^{*} = \frac{\pi}{2\varepsilon_{k_{c}}^{f}}.
\end{equation}
The critical wave vectors $k_{c}$ are obtained by solving the equation
\begin{equation}\label{critical.wave.vector.homogeneous}
  \ln{\left(\frac{|p_{k_{c}1}|^{2}}{1-|p_{k_{c}1}|^{2}}\right)}=0 \Rightarrow |p_{k_{c}1}|^{2}=1-|p_{k_{c}1}|^{2}=\frac{1}{2}.
\end{equation}
Notably, a single $k_{c}$ corresponds to a group of periodically-spaced $t_{n}^{*}$s in the complex time plane. Moreover, the condition $|p_{k_{c}1}|=1-|p_{k_{c}1}|^{2}=\frac{1}{2}$ for the existence of $k_c$ is also crucial to the occurrence of DQPTs in the homogeneous case \footnote{It is easy to check that at the critical wave vector $k_{c}$ and the first critical time $t_{0}^{*}$, the initial state is given by $\mathrm{|\psi_{0}\rangle=p_{k_{c}1}|u_{k_{c}1}^{f}\rangle+p_{k_{c}2}|u_{k_{c}2}^{f}\rangle}$, and the time-evolved state is
$\mathrm{|\psi(t)\rangle=i(p_{k_{c}1}|u_{k_{c}1}^{f}\rangle-p_{k_{c}2}|u_{k_{c}2}^{f}\rangle)}$. Therefore, it is necessary that  $\mathrm{\langle\psi_{0}|\psi(t)\rangle=i(|p_{k_{c}1}|^{2}-|p_{k_{c}2}|^{2})=0}$ in order to satisfy the condition of the DQPT.}.

For more complicated multi-band systems with $l>2$, the Fisher zeros method becomes less practical since it becomes challenging to solve the equation
\begin{equation}\label{dqpt.condition}
  \mathcal{G}(k_{cn}, t_{n}^{*}) = \sum_{\nu=1}^{l}|p_{k_{cn}\nu}|^{2}e^{-i\varepsilon_{k_{cn}\nu}^{f}t_{cn}} = 0.
\end{equation}
In addition, a single critical wave vector only corresponds to a single critical time in this case, which is different from the aforementioned homogeneous systems. This will become clearer in our later discussions.

According to Eq.(\ref{LA.polar}), the Loschmidt amplitude $\mathcal{G}_k(t)$ vanishes at the critical wave vector and critical times, leading to the ill-definedness of the phase. Therefore, the singular behavior of $\phi_k$ provides an alternative way to identify DQPTs. Note $\phi_k(t)$ is the total relative phase between the time-evolved and the initial states, which contains two components, the dynamical phase and the so-called Pancharatnam geometric phase (PGP) \cite{Berry1984392, Samuel198860}. The dynamical phase, given by
\begin{equation}\label{dyn.phase}
  \phi_{k}^{dyn}(t) = -\int_{0}^{t}ds\langle\psi(s)|H_{k}^{l}(h_{1})|\psi(s)\rangle = \sum_{\nu=1}^{l}|p_{k\nu}|^{2}\varepsilon_{k\nu}^{f}t,
\end{equation}
is proportional to $t$, ensuring that it is manifestly analytical. Therefore, the nonanalytic behavior can only come from the PGP $\phi_{k}^{G}(t)$, which is given by
\begin{equation}\label{PGP.Bloch}
  \phi_{k}^{G}(t) = \phi_{k}(t) - \phi_{k}^{dyn}(t).
\end{equation}
The behavior of PGP can be visualized by plotting it in the ($k,t$) plane. In this representation, the singularities of PGP appear as ``dynamical vortices'' located at the critical times and critical wave vectors \cite{Budich201693, Vogel201714}. Additionally, one can introduce the winding number to measure the accumulation of PGP when evolving in the Brillouin zone. Many studies have shown that the winding number is quantized and exhibits discrete jumps when moving between adjacent critical times, making it a useful dynamical topological order parameter to classify different types of DQPTs \cite{Dutta201796, Lang201898, Qiu201898, Zhou201898, Jafari2021103, Jafari2022105}.

For multi-band systems, there is another useful criterion to predict the occurrence of DQPTs \cite{Huang2016117}, which was first applied in the real Hofstadter model \cite{Huang2016117}. Explicitly, it requires
\begin{eqnarray}
  \psi_{\text{MaxMin}} &\equiv & \max_{\nu}[\min_{k}|p_{k\nu}|], \label{maxmin} \\
  \psi_{\text{MaxMin}} &=& 0 \Leftrightarrow \text{DQPT}, \label{criterion.DQPT}
\end{eqnarray}
where $\psi_{\text{MaxMin}}$ is defined as the maximum value of the minimum transition coefficients subject to all wave vectors and energy bands of the system. In the rest of the paper, we will apply this criterion as well as the singular behaviors of the rate function and PGP to study DQPTs of the multi-band complex Bloch Hamiltonian. Their effectiveness can be mutually confirmed and will provide deep insights into the origin of the aperiodic distribution of critical times.


\begin{figure}
  \centering
  \includegraphics[width=0.97\linewidth]{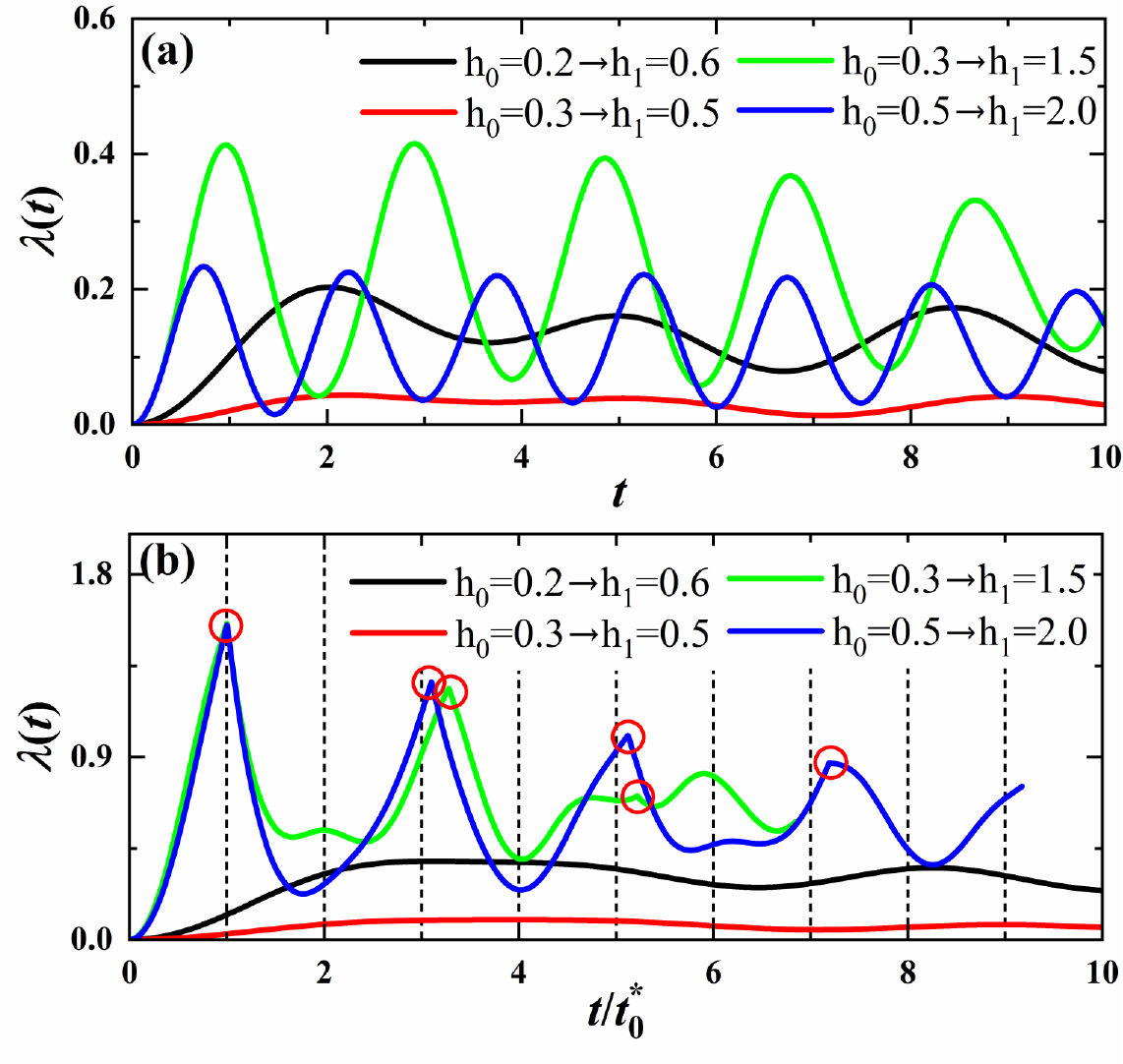}\\
  \caption{The rate functions resulting from the quench protocols initiated (a) from $|u_{k1}^{i}\rangle$, and (b) from $|u_{k2}^{i}\rangle$. The black and red lines correspond to the quenches without crossing the critical point $h_{c}\approx0.7071$, and the green and blue lines correspond to those crossing $h_{c}$. In  panel (b), the cusp-like singularities are circled and 
  the time axis is scaled by the first critical time $t_{0}^{*}$ to emphasize the nonuniformly spaced critical times.
  }\label{DQPT.l=4}
\end{figure}


\section{Behaviors of DQPT}

As an explicit example, we focus on the dynamical behavior of the period-two Bloch Hamiltonian $H_{k}^{l=4}(h)$ after a sudden quench. To facilitate comparisons between different initial states, we first examine the rate functions of the Loschmidt echo for quenches from Bloch states $|u_{k2}^{i}\rangle$ and $|u_{k1}^{i}\rangle$. It is worth noting that quenches from $|u_{k3}^{i}\rangle$ and $|u_{k4}^{i}\rangle$ are essentially the same as those from $|u_{k2}^{i}\rangle$ and $|u_{k1}^{i}\rangle$, respectively, due to the symmetric energy spectra about the zero energy, as depicted in Fig.~\ref{energy.spectra}~(a). This can be attributed to the particle-hole symmetry \cite{Altland199755}. Therefore, it is sufficient to discuss the quenches from $|u_{k1}^{i}\rangle$ and $|u_{k2}^{i}\rangle$ only.

Fig.~\ref{DQPT.l=4}~(a) displays the rate functions associated with four quench protocols initiated from lowest band state $|u_{k1}^{i}\rangle$. Two of them (green and blue lines) cross the critical point $h_{c}\approx0.7071$ while the other two do not. Obviously, none of these protocols induces a DQPT since no cusp-like singularities appear. As a comparison, Fig.~\ref{DQPT.l=4}~(b) shows the rate function associated with four protocols initiated from the state $|u_{k2}^{i}\rangle$, where the energy gap between the bands $\varepsilon_{k2}$ and $\varepsilon_{k3}$ closes at $h_{c}$. Apparently, the two protocols without crossing $h_{c}$ do not induce a DQPT, while the other two that cross $h_{c}$ do exhibit cusp-like singularities. The results suggest that two necessary conditions must be met for the occurrence of DQPT. Firstly, the initial state must have a closing gap with other energy bands at the critical point. Secondly, the quench must cross the critical point. Moreover, by scaling the time axis by the first critical time $t_{0}^{*}$ we find that the critical times $t_{n}^{*}$ are not integer multiples of $t_{0}^{*}$. This indicates that the critical times are not periodically spaced for the period-two Bloch Hamiltonian.

\begin{figure}
  \centering
  \includegraphics[width=1.0\linewidth]{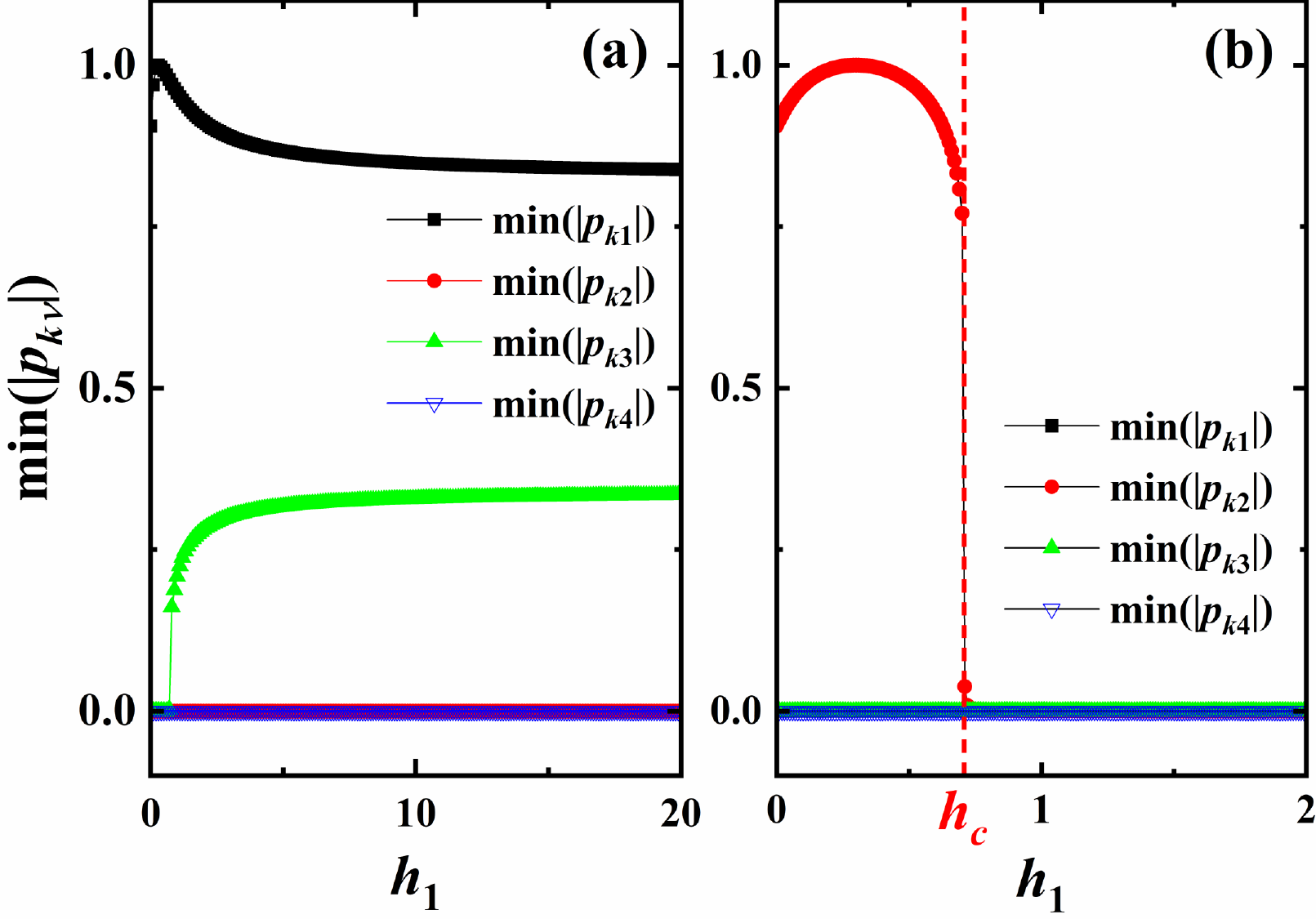}\\
  \caption{Plots of $\text{min}(|p_{k\nu}|)$ $(\nu=1,\cdots,4)$ as a function of the post-quench parameter $h_{1}$ subject to $h_0=0.3$ for a quench starting (a) from  $|u_{k1}^{i}\rangle$, and (b) from $|u_{k2}^{i}\rangle$. The DQPT can be identified by $\psi_{\text{MaxMin}}(h_{1})=\text{Max}[\text{min}(|p_{k\nu}|)]=0$ [see Eqs.~(\ref{maxmin}) and (\ref{criterion.DQPT})].
  }\label{citerion.l=4}
\end{figure}

The first necessary condition can also be justified by the criterion (\ref{criterion.DQPT}). In Fig.~\ref{citerion.l=4}~(a) and (b), we plot the $\text{min}(|p_{k\nu}|)$ $(\nu=1,\cdots,4)$ as functions of the post-quench parameter $h_{1}$ subject to $h_{0}=0.3$ for the quenches starting from  $|u_{k1}^{i}\rangle$ and $|u_{k2}^{i}\rangle$ respectively. In panel (a), $\psi_{\text{MaxMin}}(h_{1})=\text{Min}(|p_{k1}|)>0$ is always valid, thus no DQPT occurs for the quench from $|u_{k1}^{i}\rangle$. This is owing to the existing energy gap between $\varepsilon_{k1}^{i}$ and other bands, which makes it highly probable for the system to stay at the instantaneous state $|u_{k1}^{f}\rangle$, corresponding to an adiabatic evolution. Panel (b) describes the quench from $|u_{k2}^{i}\rangle$, in which $\varepsilon_{k2}$ and $\varepsilon_{k3}$ degenerate at the critical point $h_{c}$. When $h_{1}<h_{c}$, $\psi_{\text{MaxMin}}(h_{1})=\text{min}(|p_{k2}|)>0$, which is similar to the case in panel (a). However, when $h_{1}\ge h_{c}$, $\text{min}(|p_{k2}|)$ abruptly drops to zero. Therefore, all $\text{min}(|p_{k\nu}|)$ vanish in this regime, indicating the occurrence of DQPT. These results are totally consistent with those in Fig.~\ref{DQPT.l=4}.

\begin{figure}
  \centering
  \includegraphics[width=0.97\linewidth]{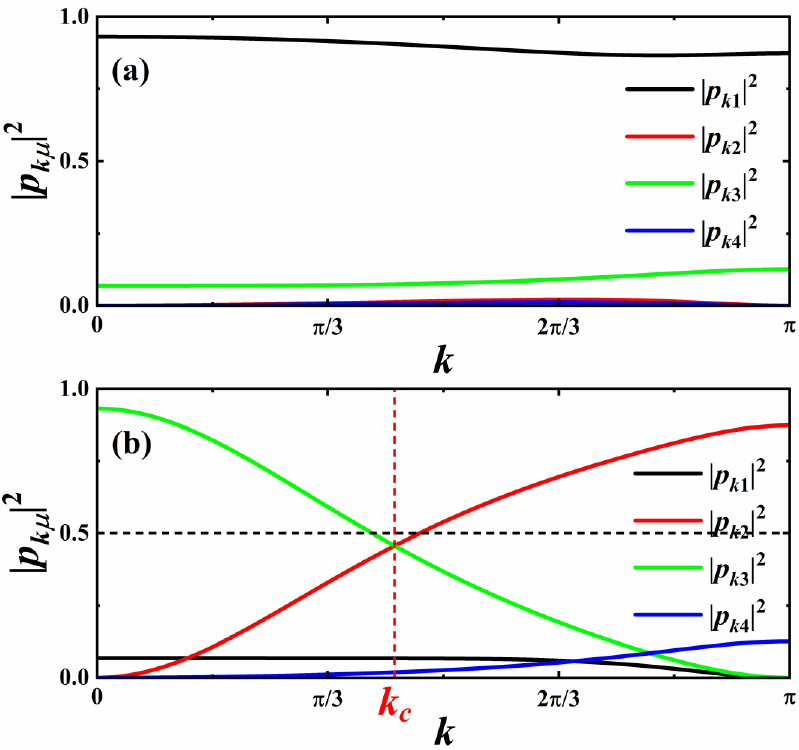}\\
  \caption{The expansion coefficients $|p_{k\nu}|^{2}=|\langle u_{k\nu}^{f}|u_{k\mu}^{i}\rangle|^{2}$ in the quenches (a) from $|u_{k1}^{i}\rangle$, and (b) from $u_{k2}^{i}$ along quench path $h_{0}=0.3\rightarrow h_{1}=1.5$. Here $k_{c}$ is the critical wave vector at which $|p_{k_c2}|^2=|p_{k_c3}|^2$, ensuring the occurrence of the first DQPT. }\label{pk.l=4}
\end{figure}

To understand the physical origin of the second necessary conditions for DQPT, we study the behaviors of the coefficients $|p_{k\nu}|^{2}=|\langle u_{k\nu}^{f}|u_{k\mu}^{i}\rangle|^{2}$ in Fig.~\ref{pk.l=4} for the two types of quench protocols discussed previously. In the top panel (a), the quench is from $|u_{k1}^{i}\rangle$ with no gap closing point, and $|p_{k\nu}|^{2}=|\langle u_{k\nu}^{f}|u_{k1}^{i}\rangle|^{2}, \nu=1,\cdots,4$ are plot for the quench path from $h_{0}=0.3$ to $h_{1}=1.5$. Obviously, the nonadiabatic transition coefficients $|p_{k2}|^{2}$, $|p_{k3}|^{2}$ and $|p_{k4}|^{2}$ are much smaller than $|p_{k1}|^{2}$. This is because the energy gap between $\varepsilon_{k1}$ and other bands keeps nonzero during the entire process. Thus, the subsequent state $|\psi_k(t)\rangle$ has a much higher probability to stay at the instantaneous eigenstate evolved from $|u_{k1}^{i}\rangle$. In other words, it is impossible for them to be perpendicular to each other according to Eq.(\ref{LA.Bloch}), which forbids the occurrence of DQPTs. On the contrary, in the bottom panel (b) we consider a quench from the state $|u_{k2}^{i}\rangle$ with a gap closing point at $h_{c}$.  It can be found that $|p_{k2}|^{2}$ and $|p_{k3}|^{2}$ intersect at the critical wave vector $k_{c}\approx1.35$ such that
\begin{equation}\label{kc.l=4}
  |\langle u_{k_{c}2}^{f}|u_{k_{c}2}^{i}\rangle|^{2} = |\langle u_{k_{c}3}^{f}|u_{k_{c}2}^{i}\rangle|^{2} < \frac{1}{2}.
\end{equation}
This indicates that the initial state $|u_{k2}^{i}\rangle$ has a large probability to transition to the state $|u_{k3}^{f}\rangle$ at $k\le k_c$ after the quench, which is reasonable since the bands $\varepsilon_{k2}$ and $\varepsilon_{k3}$ close at $h_c$. This makes it possible for the subsequent state to be perpendicular to the initial state at some critical times. The first critical time $t^*_0\approx 1.46$ can be obtained numerically from Fig.\ref{DQPT.l=4} (b).
Eq.~(\ref{critical.wave.vector.homogeneous}) suggests that the origin of $k_{c}$ in the four-band case is similar to that of the two-band models. The slight difference is that the additional transition coefficients $|p_{k1}|^{2}$ and $|p_{k4}|^{2}$ in the four-band model suppress the values of the transition probabilities in Eq.(\ref{kc.l=4}) such that $|p_{k_c2}|^{2}=|p_{k_c3}|^{2}$ are both less than $\frac{1}{2}$. This multi-band effect is in fact the main reason for the aperiodic spacing of the successive critical times, which will become clear later.

\begin{figure}
  \centering
  \includegraphics[width=0.95\linewidth]{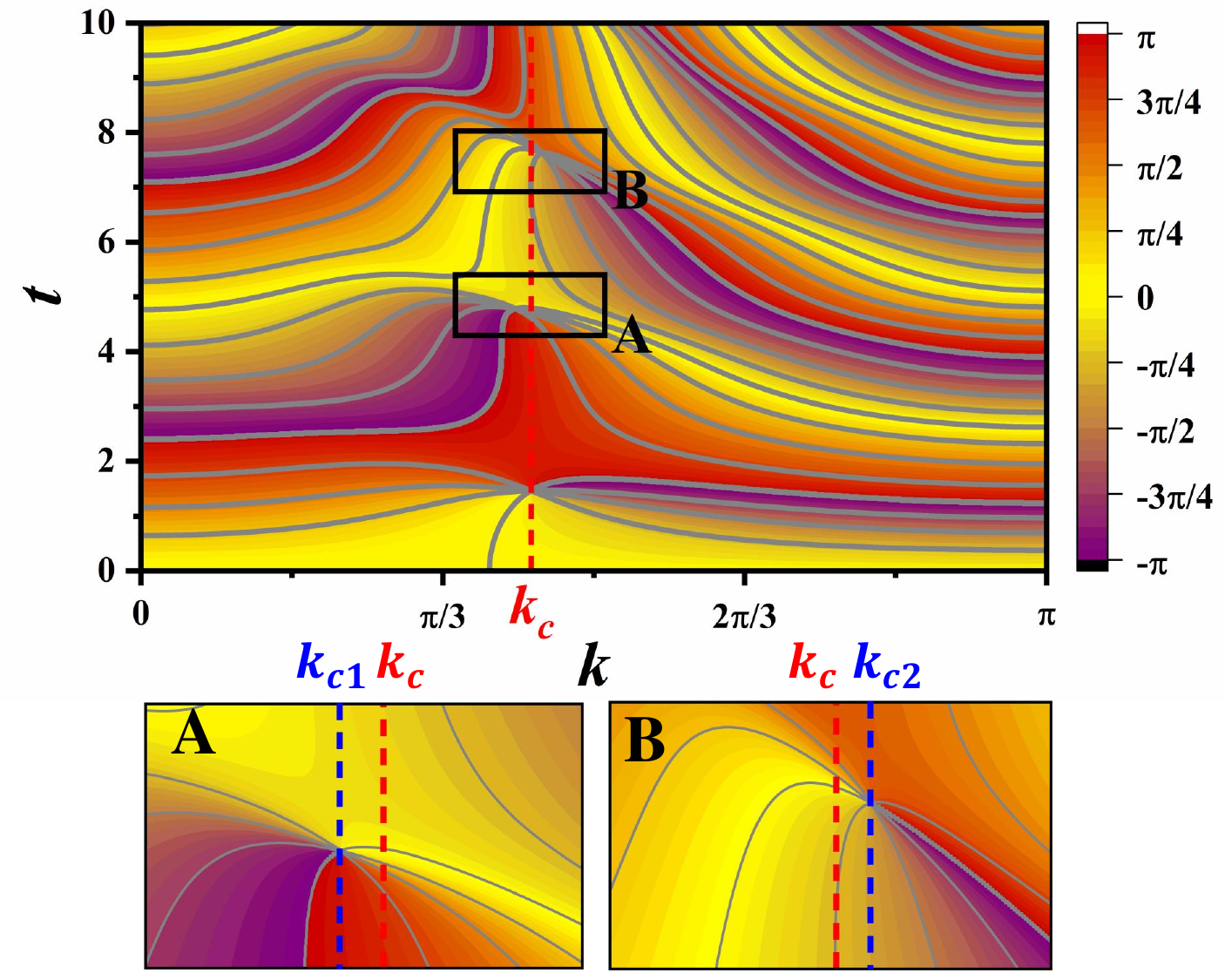}\\
  \caption{The contour plots of $\phi_{k}^{G}(t)$ in the $(k,t)$-plane for the quench starting from $|u_{k2}^{i}\rangle$ and crossing $h_{c}$ along the $h_{0}=0.3\rightarrow h_{1}=1.5$. Three dynamical vortices appear at $(k_{c}\approx1.35, t_{0}^{*}\approx1.46)$, $(k_{c1}\approx1.30, t_{1}^{*}\approx4.79)$, and $(k_{c2}\approx1.39, t_{2}^{*}\approx7.62)$, where $t^*_{0,1,2}$ are the first three critical times and $k_{c1,2}$ (marked by the blue short-dashed lines) slightly deviate from $k_c$ (marked by the red short-dashed line). The deviations are highlighted in the enlarged rectangles \textbf{A} and \textbf{B} respectively.   }\label{PGP.1=4}
\end{figure}

To investigate the subsequent DQPTs and their dependence on critical times and wave vectors, we show in Fig.~\ref{PGP.1=4} the contour plots of PGP $\phi_{k}^{G}(t)$ in the $(k,t)$ plane for quenches starting from the initial states $|u_{k2}^{i}\rangle$ and of $ h_{0}=0.3\rightarrow h_{1}=1.5$. In the top panel, there are three singular convergences of PGP, characterizing the dynamical vortices or DQPTs. The lowest one corresponds to the aforementioned first DQPT located at $(k_c,t^*_0)$. The next two DQPTs are highlighted by the rectangles \textbf{A} and \textbf{B}, which are further enlarged in the bottom panels to show more details. We emphasize that the critical wave vectors $k_{c1,2}$ are slightly deviated from $k_c$ and the associated critical times $t^*_{1,2}$ are not integral multiples of $t^*_0$. 

\begin{figure}
  \centering
  \includegraphics[width=1.0\linewidth]{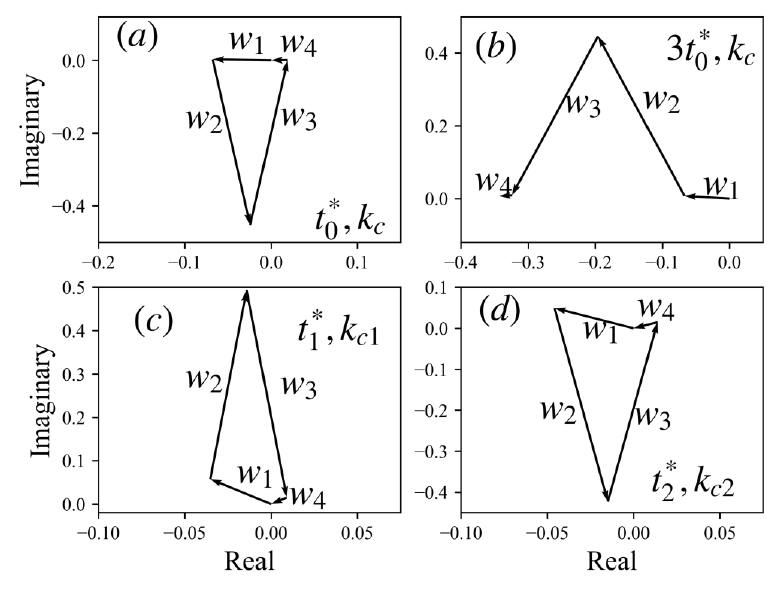}\\
  \caption{ The polygons formed by the Fisher vectors $w_{\nu}$s at (a) $(k_{c},t_{0}^{*})$, (b) $(k_{c},3t_{0}^{*})$, (c) $(k_{c1},t_{1}^{*})$, and (d) $(k_{c2},t_{2}^{*})$. 
  Note the scale of the real axis is slightly enlarged to make $w_4$ visible.  }\label{fisher.vector.l=4}
\end{figure}

To explain this discrepancy, we follow the geometrical interpolation in Refs.~ \cite{Huang2016117, Zhang_2016114}. The DQPT is specified by Eq.~(\ref{dqpt.condition}), i.e. $\mathcal{G}(k_{cn},t_{n}^{*})=0$, which is hard to solve analytically for multi-band systems. However, the above discrepancy can be understood via an intuitive picture without solving the equation. We introduce the ``Fisher vectors" $w_{\nu}=|p_{k\nu}|^{2}e^{-i\varepsilon_{k\nu}^{f}t}$ $(\nu=1,\cdots,4)$ such that Eq.~(\ref{dqpt.condition}) becomes $\sum_{\nu=1}^{4}w_{\nu}=0$, forming a closed polygon in the complex $w$-plane at DQPTs. In Fig.~\ref{fisher.vector.l=4}~(a), we show the polygon associated with the first DQPT, where $|w_2|=|w_3|=|p_{k_c2}|^{2}$ at $k_c$. Afterwards, if $w_\nu$, $\nu=1,2,3,4$ rotates with the same speed, they will form a rotating polygon identical to that in Fig.~\ref{fisher.vector.l=4}~(a). However, this is impossible since it is equivalent to the fact that the DQPT keeps occurring for $t>t^*_0$. In fact, $-\varepsilon_{k1}^{f}=\varepsilon_{k4}^{f}>-\varepsilon_{k2}^{f}=\varepsilon_{k3}^{f}$ (see Fig.\ref{energy.spectra}), which means that $w_1$ and $w_4$ rotate with the same but opposite velocities. The same result holds for $w_2$ and $w_3$ with a smaller speed. In Fig.~\ref{fisher.vector.l=4}~(b), we try to draw $w_\nu$s at $(k_c,3t^*_0)$. Obviously, they can not form a closed polygon and no DQPT occurs accordingly, which agrees with the previous assertion that later critical times are not integer multiples of $t^*_0$. Since $w_\nu$s rotate with different speeds, the closed polygon at $t>t^*_0$ must not be identical to that at $t^*_0$, indicating the edges of the polygon, i.e. the modulus of $w_\nu$s must change. Since $|w_\nu|=|p_{k\nu}|^{2}$ only depends on $k$, the critical wave vectors at later DQPTs must be different from $k_c$, which has been confirmed by the bottom panels in Fig.~\ref{PGP.1=4}. In panels (c), (d) of Fig.~\ref{fisher.vector.l=4}, we present the corresponding polygons formed by $w_\nu$s. Therefore, the aperiodic spacing of critical times is a multi-band effect since the polygon must have at least three edges, i.e. the number of bands is no less than 3.

To show the generality of our results, we did a similar study on the six-band model $H_{k}^{l=6}$. There is no significant difference between it and $H_{k}^{l=4}$, so we outline the main results in Appendix~B.

\section{DTOP in the multi-band Bloch Hamiltonian}

As mentioned in Sec.~\uppercase\expandafter {\romannumeral2}, the DTOP has been proposed to describe the topological features that emerge in DQPTs. Specifically, for two-band systems, the DTOP is evaluated by integrals over adjacent critical times and exhibits unit jumps at DQPTs, making it a useful tool to characterize the topological properties of DQPTs. However, several studies have presented some counterexamples\cite{Ding2020102, Jafari2021103, Cao2023}. In particular, Ref.~\cite{Cao2023} has reported that the periodic modulation can break the integer quantization of the DTOP in the quantum Ising chain, where the system is initiated from BCS-like ground states. In this section, we aim to investigate the influence of periodic modulation on DTOP in the Bloch Hamiltonian.

The DTOP is defined as the winding number associated with the PGP, given by \cite{Budich201693}
\begin{equation}\label{winding.number.definition}
  \nu_{D}(t) = \frac{1}{2\pi}\int_{0}^{\pi}\frac{\partial \phi_{k}^{G}(t)}{\partial k}dk,
\end{equation}
where $\phi_{k}^{G}(t)$ is obtained from Eq.~(\ref{PGP.Bloch}). Here the integral is taken  over $(0,\pi]$ instead of $(-\pi,\pi]$ due to the symmetry of $\phi_{k}^{G}(t)$ \cite{Zhou202123}. For example, in the Bloch Hamiltonian, $\phi_{k}^{G}(t)$ has an inversion symmetry with respect to $k=0$, i.e., $\phi_{k}^{G}(t)=\phi_{-k}^{G}(t)$. Hence the integral must be evaluated over the reduced Brillouin zone (BZ) $k\in(0,\pi]$ since the integral vanishes over the whole BZ $k\in(-\pi,\pi]$. Integrating $k$, Eq.(\ref{winding.number.definition}) gives 
\begin{equation}\label{winding.number.calculation}
  \nu_{D}(t) = \frac{\phi_{k=\pi}^{G}(t)-\phi_{k=0}^{G}(t)}{2\pi}+\mathcal{N},
\end{equation}
where $\frac{\phi_{k=\pi}^{G}(t)-\phi_{k=0}^{G}(t)}{2\pi}$ represents the accumulated phase difference at the boundaries of the reduced BZ, and $\mathcal{N}$ is the number of times that $\phi_{k}^{G}(t)$ is folded into its principal angle value. Specifically, $\mathcal{N}$ increases by one when folding from $\pi$ to $-\pi$, and decreases by one when folding from $-\pi$ to $\pi$. Note the PGP is ill-defined at critical times, which induces a $2\pi$ jump (folding) at DQPTs. Thus, the folding term $\mathcal{N}$ is always an integer, and experiences unit changes at critical times. However, the boundary term does not necessarily take integer values, which may make the DTOP non-integer quantized.

\begin{figure}
  \centering
  \includegraphics[width=0.9\linewidth]{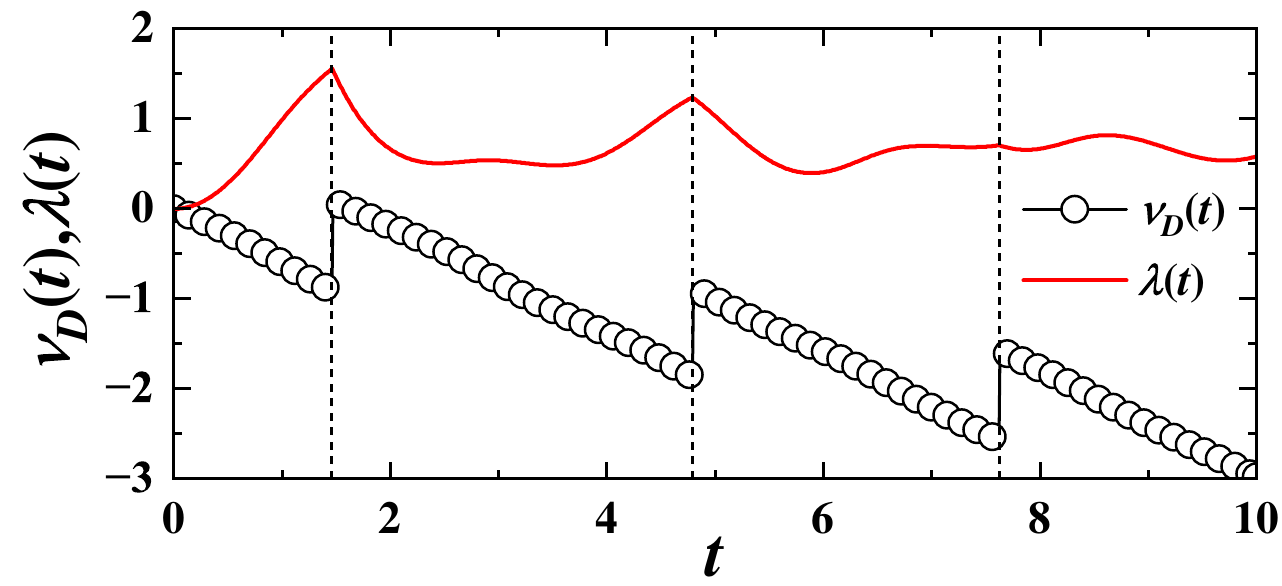}\\
  \caption{The winding numbers $\nu_{D}(t)$ for the quench from the Bloch state $|u_{k2}^{i}\rangle$ in the Bloch Hamiltonian $H_{k}^{l=4}$. The quench path is from $h_{0}=0.3$ to $h_{1}=1.5$.}\label{DTOP}
\end{figure}

Fig.~\ref{DTOP} displays the winding number as a function of $t$ for the quench from $|u_{k2}^{i}\rangle$ along the path $h_{0}=0.3\rightarrow h_{1}=1.5$. Apparently, $\nu_{D}(t)$ is no longer integer-quantized but still exhibits a unit jump at critical times. 
The reason can be deduced from Fig.~\ref{PGP.1=4}, in which we can find that $\phi_{k=0}^{G}(t)$ and $\phi_{k=\pi}^{G}(t)$ vary with different changing rates. Obviously, $\phi_{k=\pi}^{G}(t)$ changes faster than $\phi_{k=0}^{G}(t)$ such that $\frac{\phi_{k=\pi}^{G}(t)-\phi_{k=0}^{G}(t)}{2\pi}$ is not a constant. Hence, the winding number is neither quantized nor topological. Nevertheless, it still exhibits discrete jumps due to the nonanalytic nature of the PGP at critical times.


\section{Mixed-state DQPT in the multi-band Bloch Hamiltonian}

Building upon the previous discussion, we can now broaden our theory to include the mixed state and explore the impact of multiband on the mixed-state DQPT. This is a significant concern as, in practical experiments \cite{Vogel201714, Jurcevic2017119, Muniz.580.602}, the initial state prepared for the system far from equilibrium is often the naturally mixed state rather than the pure state. The concept of the generalized Loschmidt amplitude (GLA) for mixed states has been extensively studied and is now well-established \cite{Bhattacharya.96.180303, Heyl.96.180304, Lang.98.134310, Bandyopadhyay.8, Hou.102.104305, Kyaw.101.012111, Mera.97.094110, Sedlmayr.97.045147, Link.125.143602, Hou.104.023303}. Specifically, studies have shown that in two-band models, the mixed-state DQPT exhibits non-analyticities. Interestingly, the critical wave vectors associated with these non-analytic behaviors are found to be independent of both temperature and probability \cite{Bhattacharya.96.180303, Heyl.96.180304, Lang.98.134310}. 

\begin{figure}
  \centering
  \includegraphics[width=0.9\linewidth]{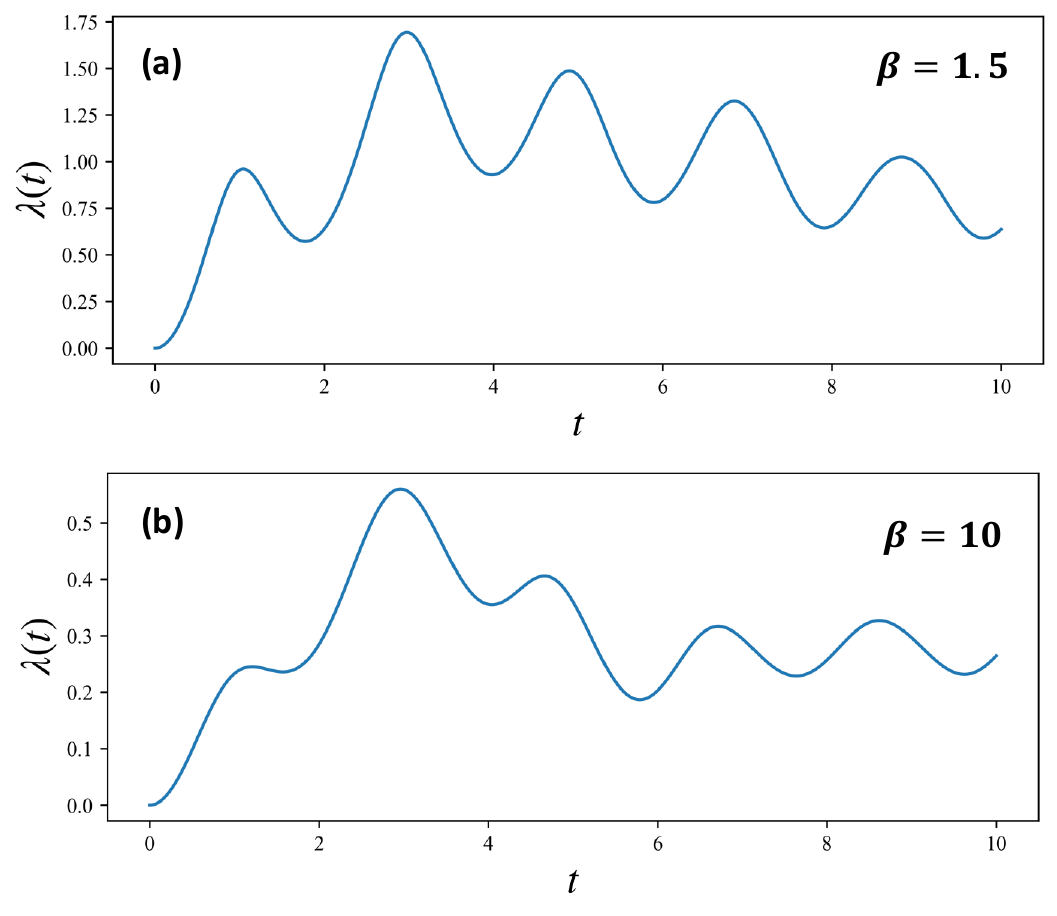}\\
  \caption{The rate functions for the mixed state with (a) $\beta=1.5$, and (b) $\beta=10$. The quench path is from $h_{0}=0.5$ to $h_{1}=1.5$.}\label{mixed}
\end{figure}

Consider an initial mixed state described by a full-ranked density matrix $\rho_{0}$. At $t=0$, a sudden quench is implemented, and the Hamiltonian of the system is quenched to $H$ such that $[\rho_{0}, H]\neq0$. After the quench, the density matrix is $\rho(t) = e^{-iHt}\rho_{0}e^{iHt} = U(t)\rho_{0}U^{\dag}(t)$, where $U(t)=e^{-iHt}$ is the time-evolution operator. The definition of the GLA is 
\begin{equation}
    \mathcal{G}L(t) = \text{Tr}[\rho_{0}U(t)] = \prod_{k>0}\mathcal{G}L_{k}(t) = \prod_{k>0}\text{Tr}[\rho_{0k}U_{k}(t)],
\end{equation}
where we decompose the GLA for every $k$ $(k>0)$. In our model $H_{k}^{l=4}$, the initial density matrix takes the form
\begin{equation}
    \rho_{0k} = \sum_{\mu=1}^{L} f_{k\mu}|u_{k\mu}^{i}\rangle\langle u_{k\mu}^{i}|,
\end{equation}
where the probabilities $f_{k\mu}\in(0,1]$ $(\mu=1,\cdots,L)$ of the electron being in $|u_{k\mu}^{i}\rangle$ parameterize the translation-invariant generalized Gibbs state. According to Eq.~\ref{time-evolved}, by selecting the eigenstates of the post-quench Hamiltonian as the basis, we obtain the expression of GLA as
\begin{equation}\label{GLA.l=4}
    \mathcal{G}L_{k}(t) = \sum_{\mu=1}^{L}f_{k\mu}\sum_{\nu=1}^{L}|\langle u_{k\nu}^{f}|u_{k\mu}^{i}\rangle|^{2}e^{-i\varepsilon_{k\nu}^{f}t}.
\end{equation}
The occurrence of DQPT demands that the GLA vanishes, i.e. $\mathcal{G}L_{k}(t) = 0$. However, this will never happen for the mixed state according to Eq.~(\ref{GLA.l=4}). The components $f_{k1}\sum_{\nu=1}^{L}|\langle u_{k\nu}^{f}|u_{k1}^{i}\rangle|^{2}e^{-i\varepsilon_{k\nu}^{f}t}$ and $f_{k4}\sum_{\nu=4}^{L}|\langle u_{k\nu}^{f}|u_{k4}^{i}\rangle|^{2}e^{-i\varepsilon_{k\nu}^{f}t}$ are actually corresponding to the Loschmidt amplitude from $|u_{k1}^{i}\rangle$ and $|u_{k4}^{i}\rangle$, respectively. The previous investigation in Section.~\uppercase\expandafter {\romannumeral3} already shows that the DQPT can not appear in the quenches from $|u_{k1}^{i}\rangle$ and $|u_{k4}^{i}\rangle$. These two components thus can never equal zero. While for the components $f_{k2}\sum_{\nu=1}^{L}|\langle u_{k\nu}^{f}|u_{k2}^{i}\rangle|^{2}e^{-i\varepsilon_{k\nu}^{f}t}$ and $f_{k3}\sum_{\nu=4}^{L}|\langle u_{k\nu}^{f}|u_{k3}^{i}\rangle|^{2}e^{-i\varepsilon_{k\nu}^{f}t}$, they only equal zero at the certain critical times and critical wave vectors of pure state DQPTs. Therefore, it is impossible for the $\mathcal{G}L_{k}(t)$ to satisfy the condition for the occurrence of DQPT. Fig.~\ref{mixed} displays the rate functions for the mixed state with (a) $\beta=1.5$ and (b) $\beta=10$. In both cases, the rate functions are smooth curves, indicating the absence of DQPTs. This characteristic differs significantly from the behavior observed in the two-band model, highlighting the distinct influence of multiple bands on the system.

\section{Conclusion}

We investigate the occurrence of DQPTs in the multi-band Bloch systems after a quench from different types of Bloch bands. It is found that DQPTs strongly depend on the initial states, and the non-adiabatic evolution of the energy band with a gap closing point also plays a crucial role. In contrast to two-band systems, the appearance of DQPTs in multiple-band systems requires two necessary conditions. Firstly, the pre-quench initial state must degenerate with another band at a certain critical point. Secondly, the quantum quench must cross that critical point. Thus, the system can evolve to a state orthogonal to the initial state after the quench. In addition, we have identified the multi-band effect as the key factor contributing to the non-uniform distribution of critical times, which also has an interesting geometric interpretation with the help of Fisher vectors. Furthermore, we examined the impact of the multi-band effect on the DTOP and found that the winding number is no longer quantized and topological. However, its discrete jumps can still be applied to characterize DQPTs. In addition, we study the influence of multiple bands on the mixed-state DQPT and find that the DQPT is absent at non-zero temperatures. The reason can be attributed to the absence of pure state DQPT in the quench from the state corresponding to the gapped band.

\begin{acknowledgments}
  K. Cao acknowledges Professor Peiqing Tong for extensive discussions and critical comments on the manuscript.
  The work is supported by the National Key Basic Research Program of China (No.~2020YFB0204800), the National Natural Science Foundation of China (No.~12074064), and Key Research Projects of Zhejiang Lab (Nos. 2021PB0AC01 and 2021PB0AC02).
\end{acknowledgments}

\appendix

\section{Bloch Hamiltonian based on the periodic Kitaev model}

We study the Hamiltonian that describes particles in a one-dimensional lattice subjecting to periodic effects, based on the one-dimensional Kitaev model
\cite{Cao202236}
\begin{equation}\label{kitaev}
  H = -\frac{1}{2}\sum_{n=1}^{N}\{[J_{n}c_{n}^{\dag}c_{n+1}+\Delta_{n}c_{n}^{\dag}c_{n+1}^{\dag}+hc_{n}^{\dag}c_{n}]+h.c.\},
\end{equation}
where $J_{n}$ are hopping interactions, $\Delta_{n}$ are superconducting gaps, and $h$ is the external field. We take $\Delta_{n}=J_{n}$ in our work for simplicity.

Under the periodic boundary condition, we can express the Hamiltonian of the system as the form
\begin{equation}\label{Hamil.k}
  H=\sum_{k>0}\Psi^{\dag}_{k}H_{k}\Psi_{k}
\end{equation}
in the momentum space $k>0$, where the spinor operator is $\Psi^{\dag}_{k}=(c^{\dag}_{k1},c_{-k1},\cdots,c^{\dag}_{kL},c_{-kL})$ and $H_{k}$ is the associated Bloch Hamiltonian.

Specifically, for the period-two case, we consider the nearest-neighbor interactions:
\begin{equation}\label{two_J}
  J_{n}=\left\{\begin{array}{cr}
                 J_{1}, & \text{odd}\quad n, \\
                 J_{2}, & \text{even}\quad n.
               \end{array}
  \right.
\end{equation}
Here, we set $\alpha=J_{2}/J_{1}$ and $J_{1}=J=1$ without losing generality, so that $\alpha$ denotes the strength of the periodic modulation. $\alpha=1$ recovers the homogeneous case. The Bloch Hamiltonian of the period-two case is a $4\times4$ Hermitian matrix yielding
\begin{widetext}
\begin{equation}\label{four.Bloch.Hamiltonian}
  H_{k}^{l=4}(h,\alpha) = \frac{J}{2}
  \left(\begin{array}{cccc}
                -2h/J & 0 & -(1+\alpha e^{-ik}) & -(1-\alpha e^{-ik}) \\
                0 & 2h/J & (1-\alpha e^{-ik}) & (1+\alpha e^{-ik}) \\
                -(1+\alpha e^{ik}) & (1-\alpha e^{ik}) & -2h/J & 0 \\
                -(1-\alpha e^{ik}) & (1+\alpha e^{ik}) & 0 & 2h/J \\
  \end{array}\right),
\end{equation}
\end{widetext}
where $l=4$ denotes that the Bloch Hamiltonian has four Bloch bands.

Similarly, for the period-three three cases, we consider the nearest-neighbor interactions ($p\in\mathbb{Z}$)
\begin{equation}\label{three.J}
  J_{n}=\left\{\begin{array}{cl}
                   J, & n=3p-2, \\
                   \alpha J, & n=3p-1,   \\
                   \beta J, & n=3p.
                 \end{array}
    \right.
\end{equation}
For simplicity, we set $\beta=1$ and use $\alpha$ to control the strength of the periodic modulation. The Bloch Hamiltonian of the period-thee case is a $6\times6$ Hermitian matrix yielding
\begin{widetext}
\begin{equation}\label{Hk_three}
  H_{k}^{l=6}(h,\alpha)=\frac{J}{2}\left(
            \begin{array}{cccccc}
              -2h/J    & 0       & -1  & -1 & -e^{-ik} & e^{-ik} \\
              0 & 2h/J & 1  & 1 & -e^{-ik} & e^{-ik} \\
              -1 & 1 & -2h/J & 0 & -\alpha & -\alpha \\
              -1 & 1 & 0 & 2h/J & \alpha & \alpha \\
              -e^{ik}  & -e^{ik} & -\alpha & \alpha & -2h/J & 0 \\
              e^{ik}   & e^{ik}  & -\alpha & \alpha & 0 & 2h/J \\
            \end{array}
          \right),
\end{equation}
\end{widetext}
where $l=6$ denotes that the Bloch Hamiltonian has six Bloch bands. For $H_{k}^{l=6}(h)$, the energy gap between $\varepsilon_{k3}$ and $\varepsilon_{k4}$ vanishes at $h_{c}=\sqrt[3]{\alpha}$.

\section{DQPT in six band Hamiltonian $H_{k}^{l=6}$}

\begin{figure}
  \centering
  \includegraphics[width=1.0\linewidth]{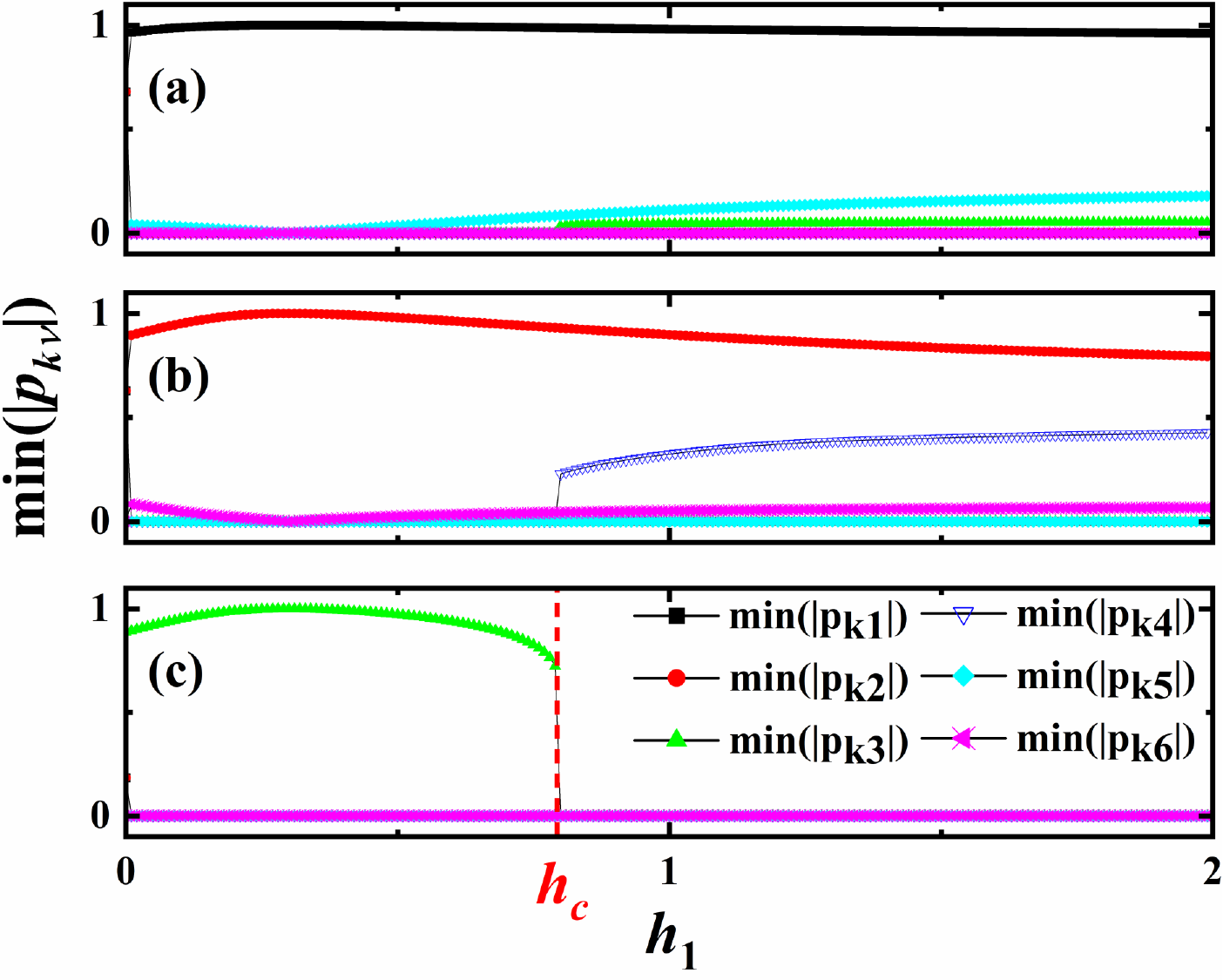}\\
  \caption{Plots of $\text{min}(|p_{k\nu}|)$ $(\nu=1,\cdots,6)$ of  $H_{k}^{l=6}(h)$ as a function of $h_{1}$ subject to $h_{0}=0.3$ for the quench (a) from $|u_{k1}^{i}\rangle$, (b) from $|u_{k2}^{i}\rangle$, and (c) from $|u_{k3}^{i}\rangle$. The occurrence of DQPTs are identified by $\psi_{\text{MaxMin}}(h_{1})=\text{Max}[\text{min}(|p_{k\nu}|)]=0$.}\label{citerion.l=6}
\end{figure}

\begin{figure}
  \centering
  \includegraphics[width=1.0\linewidth]{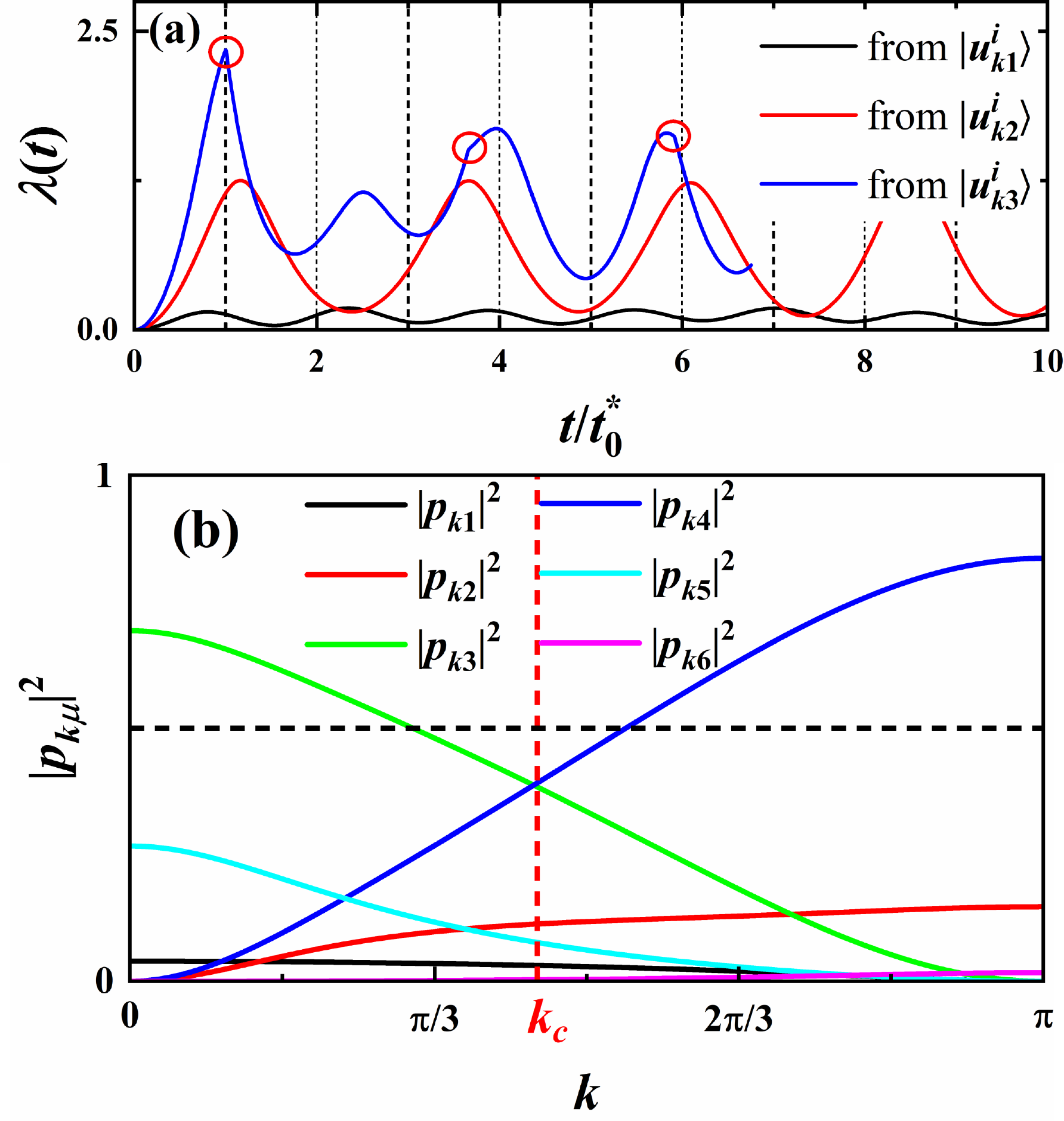}\\
  \caption{(a) The rate functions of $H_{k}^{l=6}(h)$ after quenches from different Bloch states along the path $h_{0}=0.3\rightarrow h_{1}=1.5$ that crosses $h_{c}\approx0.7937$. The time axis is scaled by  $t_{0}^{*}$ to highlight the non-uniform distribution of critical times. (b) The expansion coefficients $|p_{k\nu}|^{2}=|\langle u_{k\nu}^{f}|u_{k\mu}^{i}\rangle|^{2}$, $\nu=1,2,3$ for the quench from $|u_{k3}^{i}\rangle$ along the path $h_{0}=0.3\rightarrow h_{1}=1.5$. }\label{rate.function.l=6}
\end{figure}

The results of $H_{k}^{l=4}(h)$ in the main text can be extended to other multi-band systems like $H_{k}^{l=6}(h)$, and we briefly outline the discussions here.

In the end of Sec.~\uppercase\expandafter {\romannumeral3}~B, we have confirmed that the criterion (\ref{criterion.DQPT}) can also predict DQPTs for $H_{k}^{l=6}(h)$. Similarly, in Fig.~\ref{citerion.l=6} (a)-(c) we show $\text{min}(|p_{k\nu}|)$ $(\nu=1,\cdots,6)$ for $H_{k}^{l=6}(h)$ of $\alpha=0.5$ as a function of $h_{1}$ subject to $h_{0}=0.3$ for quenches from $|u_{k1}^{i}\rangle$, $|u_{k2}^{i}\rangle$, and $|u_{k3}^{i}\rangle$ respectively. The gapless point between the third and fourth energy bands is $h_{c}=\sqrt[3]{\alpha}\approx0.7937$. Similarly, in panels (a) and (b), we observe that $\psi_{\text{MaxMin}}(h_{1})=\text{min}(|p_{k1}|)=\text{min}(|p_{k2}|)$, which are always positive for quenches from $|u_{k1}^{i}\rangle$ and $|u_{k2}^{i}\rangle$. In panel (c), $\psi_{\text{MaxMin}}(h_{1})=\text{min}(|p_{k3}|)$ before $h_{1}$ crosses $h_{c}$ for the quench from $|u_{k3}^{i}\rangle$. This indicates that no DQPTs will occur if a quench is from a state without a gapless point or from a state with a gapless point but along a path without crossing $h_{c}$. For $h_{1}>h_{c}$, $\text{min}(|p_{k2}|)$ exhibits an abrupt change from a finite value to zero, and the criterion is thus satisfied. Similar to the case of $H_{k}^{l=4}(h)$, DQPTs only occur for quenches from $|u_{k3}^{i}\rangle$ and crossing $h_{c}$.

\begin{figure}
  \centering
  \includegraphics[width=1.0\linewidth]{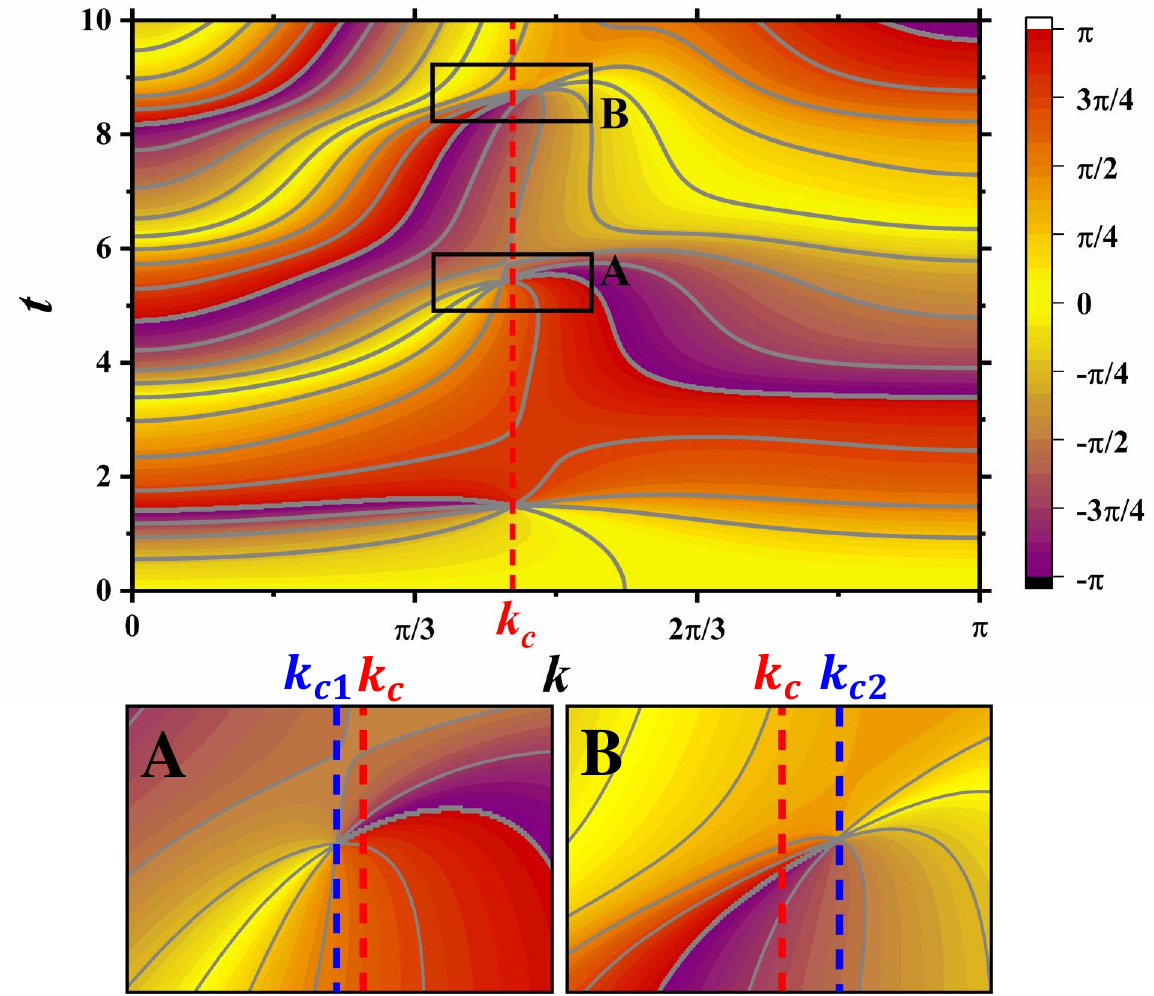}\\
  \caption{   The contour plots of $\phi_{k}^{G}(t)$ for $H_{k}^{l=6}(h)$ in the $(k,t)$-plane after the quench starting from $|u_{k3}^{i}\rangle$ and crossing $h_{c}$ along the $h_{0}=0.3\rightarrow h_{1}=1.5$. The first three dynamical vortices appear in this regime. Similarly, the details near the second and third vortices are enlarged in \textbf{A} and \textbf{B}. }\label{PGP.6}
\end{figure}

To verify the findings obtained through the analysis of $\psi_{\text{MaxMin}}(h_{1})$, we study a representative example. Fig.~\ref{rate.function.l=6}~(a) displays the rate functions for quenches from the states $|u_{k1}^{i}\rangle$, $|u_{k2}^{i}\rangle$, and $|u_{k3}^{i}\rangle$ respectively along the path $h_{0}=0.3\rightarrow h_{1}=1.5$. When quenching from $|u_{k1}^{i}\rangle$ and $|u_{k2}^{i}\rangle$, the rate functions are smooth with respect to $t$ [marked by the black and red lines in Fig.~\ref{rate.function.l=6}~(a)], indicating the absence of DQPTs. When quenching from $|u_{k3}^{i}\rangle$, the rate function exhibits cusp-like singularities at critical times which are not equidistant [marked by the blue line in Fig.~\ref{rate.function.l=6}~(a)]. These observations provide further evidence for our previous findings inferred from $\psi_{\text{MaxMin}}(h_{1})$.

In Fig.~\ref{rate.function.l=6}~(b), we present the expansion coefficients $|p_{k\nu}|^{2}=|\langle u_{k\nu}^{f}|u_{k\mu}^{i}\rangle|^{2}$, $\nu=1,2,3$ for the quench from $|u_{k3}^{i}\rangle$. It is evident that the critical momentum $k_{c}$ can be obtained from the relation:
\begin{equation}\label{kc.l=6}
  |\langle u_{k_{c}3}^{f}|u_{k_{c}3}^{i}\rangle|^{2} = |\langle u_{k_{c}4}^{f}|u_{k_{c}3}^{i}\rangle|^{2} < \frac{1}{2},
\end{equation}
which is similar to the condition (\ref{critical.wave.vector.homogeneous}) of the two-band model and has been
been tested extensively by numerical methods.
Similarly, it is found $|p_{k3}|^{2}=|p_{k4}|^{2}<\frac{1}{2}$ due to the non-adiabatic terms $|p_{k1}|^{2}$, $|p_{k2}|^{2}$, $|p_{k5}|^{2}$ and $|p_{k6}|^{2}$. At later critical times $t_{n}^{*}$ $(n>0)$, the critical momentum will not satisfy Eq.(\ref{kc.l=6}) due to the multi-band effect, analogous to the four-band model in the main text.

To understand the origin of the non-uniformly spaced critical times, we present in Fig.~\ref{PGP.6} the contour plots of $\phi_{k}^{G}(t)$ in the $(k,t)$-plane for the quench starting from $|u_{k3}^{i}\rangle$ and crossing $h_{c}$ along the path $h_{0}=0.3\rightarrow h_{1}=1.5$. Similar to the four-band model, we find that the critical momenta where the later DQPTs occur at $t_{n}^{*}$ $(n>0)$ also slightly deviate from $k_{c}$, which can be explained by the geometric interpretation just as that of $H_{k}^{l=4}(h)$ too. Therefore, the aperiodic distribution of critical times indeed comes from the multi-band effect. 

\bibliography{dqpt}

\providecommand{\noopsort}[1]{}\providecommand{\singleletter}[1]{#1}%
\begin{thebibliography}{105}%
\makeatletter
\providecommand \@ifxundefined [1]{%
 \@ifx{#1\undefined}
}%
\providecommand \@ifnum [1]{%
 \ifnum #1\expandafter \@firstoftwo
 \else \expandafter \@secondoftwo
 \fi
}%
\providecommand \@ifx [1]{%
 \ifx #1\expandafter \@firstoftwo
 \else \expandafter \@secondoftwo
 \fi
}%
\providecommand \natexlab [1]{#1}%
\providecommand \enquote  [1]{``#1''}%
\providecommand \bibnamefont  [1]{#1}%
\providecommand \bibfnamefont [1]{#1}%
\providecommand \citenamefont [1]{#1}%
\providecommand \href@noop [0]{\@secondoftwo}%
\providecommand \href [0]{\begingroup \@sanitize@url \@href}%
\providecommand \@href[1]{\@@startlink{#1}\@@href}%
\providecommand \@@href[1]{\endgroup#1\@@endlink}%
\providecommand \@sanitize@url [0]{\catcode `\\12\catcode `\$12\catcode
  `\&12\catcode `\#12\catcode `\^12\catcode `\_12\catcode `\%12\relax}%
\providecommand \@@startlink[1]{}%
\providecommand \@@endlink[0]{}%
\providecommand \url  [0]{\begingroup\@sanitize@url \@url }%
\providecommand \@url [1]{\endgroup\@href {#1}{\urlprefix }}%
\providecommand \urlprefix  [0]{URL }%
\providecommand \Eprint [0]{\href }%
\providecommand \doibase [0]{http://dx.doi.org/}%
\providecommand \selectlanguage [0]{\@gobble}%
\providecommand \bibinfo  [0]{\@secondoftwo}%
\providecommand \bibfield  [0]{\@secondoftwo}%
\providecommand \translation [1]{[#1]}%
\providecommand \BibitemOpen [0]{}%
\providecommand \bibitemStop [0]{}%
\providecommand \bibitemNoStop [0]{.\EOS\space}%
\providecommand \EOS [0]{\spacefactor3000\relax}%
\providecommand \BibitemShut  [1]{\csname bibitem#1\endcsname}%
\let\auto@bib@innerbib\@empty
\bibitem [{\citenamefont {Bloch}\ \emph {et~al.}(2008)\citenamefont {Bloch},
  \citenamefont {Dalibard},\ and\ \citenamefont {Zwerger}}]{Bloch2008}%
  \BibitemOpen
  \bibfield  {author} {\bibinfo {author} {\bibfnamefont {I.}~\bibnamefont
  {Bloch}}, \bibinfo {author} {\bibfnamefont {J.}~\bibnamefont {Dalibard}}, \
  and\ \bibinfo {author} {\bibfnamefont {W.}~\bibnamefont {Zwerger}},\ }\href
  {\doibase 10.1103/RevModPhys.80.885} {\bibfield  {journal} {\bibinfo
  {journal} {Rev. Mod. Phys.}\ }\textbf {\bibinfo {volume} {80}},\ \bibinfo
  {pages} {885} (\bibinfo {year} {2008})}\BibitemShut {NoStop}%
\bibitem [{\citenamefont {Lewenstein}\ \emph {et~al.}(2012)\citenamefont
  {Lewenstein}, \citenamefont {Sanpera},\ and\ \citenamefont
  {Ahufinger}}]{Lewenstein2012}%
  \BibitemOpen
  \bibfield  {author} {\bibinfo {author} {\bibfnamefont {M.}~\bibnamefont
  {Lewenstein}}, \bibinfo {author} {\bibfnamefont {A.}~\bibnamefont {Sanpera}},
  \ and\ \bibinfo {author} {\bibfnamefont {V.}~\bibnamefont {Ahufinger}},\
  }\href {\doibase 10.1093/acprof:oso/9780199573127.001.0001} {\emph {\bibinfo
  {title} {{Ultracold Atoms in Optical Lattices: Simulating quantum many-body
  systems}}}}\ (\bibinfo  {publisher} {Oxford University Press},\ \bibinfo
  {year} {2012})\BibitemShut {NoStop}%
\bibitem [{\citenamefont {Belsley}(2013)}]{Belsley2013}%
  \BibitemOpen
  \bibfield  {author} {\bibinfo {author} {\bibfnamefont {M.}~\bibnamefont
  {Belsley}},\ }\href {\doibase 10.1080/00107514.2013.800135} {\bibfield
  {journal} {\bibinfo  {journal} {Contemporary Physics}\ }\textbf {\bibinfo
  {volume} {54}},\ \bibinfo {pages} {112} (\bibinfo {year} {2013})}\BibitemShut
  {NoStop}%
\bibitem [{\citenamefont {Polkovnikov}\ \emph {et~al.}(2011)\citenamefont
  {Polkovnikov}, \citenamefont {Sengupta}, \citenamefont {Silva},\ and\
  \citenamefont {Vengalattore}}]{Polkovnikov2011}%
  \BibitemOpen
  \bibfield  {author} {\bibinfo {author} {\bibfnamefont {A.}~\bibnamefont
  {Polkovnikov}}, \bibinfo {author} {\bibfnamefont {K.}~\bibnamefont
  {Sengupta}}, \bibinfo {author} {\bibfnamefont {A.}~\bibnamefont {Silva}}, \
  and\ \bibinfo {author} {\bibfnamefont {M.}~\bibnamefont {Vengalattore}},\
  }\href {\doibase 10.1103/RevModPhys.83.863} {\bibfield  {journal} {\bibinfo
  {journal} {Rev. Mod. Phys.}\ }\textbf {\bibinfo {volume} {83}},\ \bibinfo
  {pages} {863} (\bibinfo {year} {2011})}\BibitemShut {NoStop}%
\bibitem [{\citenamefont {Mitra}(2018)}]{Aditi2018}%
  \BibitemOpen
  \bibfield  {author} {\bibinfo {author} {\bibfnamefont {A.}~\bibnamefont
  {Mitra}},\ }\href {\doibase 10.1146/annurev-conmatphys-031016-025451}
  {\bibfield  {journal} {\bibinfo  {journal} {Annual Review of Condensed Matter
  Physics}\ }\textbf {\bibinfo {volume} {9}},\ \bibinfo {pages} {245} (\bibinfo
  {year} {2018})}\BibitemShut {NoStop}%
\bibitem [{\citenamefont {Zurek}\ \emph {et~al.}(2005)\citenamefont {Zurek},
  \citenamefont {Dorner},\ and\ \citenamefont {Zoller}}]{Zurek2005}%
  \BibitemOpen
  \bibfield  {author} {\bibinfo {author} {\bibfnamefont {W.~H.}\ \bibnamefont
  {Zurek}}, \bibinfo {author} {\bibfnamefont {U.}~\bibnamefont {Dorner}}, \
  and\ \bibinfo {author} {\bibfnamefont {P.}~\bibnamefont {Zoller}},\ }\href
  {\doibase 10.1103/PhysRevLett.95.105701} {\bibfield  {journal} {\bibinfo
  {journal} {Phys. Rev. Lett.}\ }\textbf {\bibinfo {volume} {95}},\ \bibinfo
  {pages} {105701} (\bibinfo {year} {2005})}\BibitemShut {NoStop}%
\bibitem [{\citenamefont {Heyl}\ \emph {et~al.}(2013)\citenamefont {Heyl},
  \citenamefont {Polkovnikov},\ and\ \citenamefont {Kehrein}}]{Heyl2013110}%
  \BibitemOpen
  \bibfield  {author} {\bibinfo {author} {\bibfnamefont {M.}~\bibnamefont
  {Heyl}}, \bibinfo {author} {\bibfnamefont {A.}~\bibnamefont {Polkovnikov}}, \
  and\ \bibinfo {author} {\bibfnamefont {S.}~\bibnamefont {Kehrein}},\ }\href
  {\doibase 10.1103/PhysRevLett.110.135704} {\bibfield  {journal} {\bibinfo
  {journal} {Phys. Rev. Lett.}\ }\textbf {\bibinfo {volume} {110}},\ \bibinfo
  {pages} {135704} (\bibinfo {year} {2013})}\BibitemShut {NoStop}%
\bibitem [{\citenamefont {Zvyagin}(2016)}]{Zvyagin201642}%
  \BibitemOpen
  \bibfield  {author} {\bibinfo {author} {\bibfnamefont {A.~A.}\ \bibnamefont
  {Zvyagin}},\ }\href {\doibase 10.1063/1.4969869} {\bibfield  {journal}
  {\bibinfo  {journal} {Low Temperature Physics}\ }\textbf {\bibinfo {volume}
  {42}},\ \bibinfo {pages} {971} (\bibinfo {year} {2016})}\BibitemShut
  {NoStop}%
\bibitem [{\citenamefont {Heyl}(2018)}]{Heyl201881}%
  \BibitemOpen
  \bibfield  {author} {\bibinfo {author} {\bibfnamefont {M.}~\bibnamefont
  {Heyl}},\ }\href {\doibase 10.1088/1361-6633/aaaf9a} {\bibfield  {journal}
  {\bibinfo  {journal} {Reports on Progress in Physics}\ }\textbf {\bibinfo
  {volume} {81}},\ \bibinfo {pages} {054001} (\bibinfo {year}
  {2018})}\BibitemShut {NoStop}%
\bibitem [{\citenamefont {Vajna}\ and\ \citenamefont
  {D\'ora}(2014)}]{Vajna201489}%
  \BibitemOpen
  \bibfield  {author} {\bibinfo {author} {\bibfnamefont {S.}~\bibnamefont
  {Vajna}}\ and\ \bibinfo {author} {\bibfnamefont {B.}~\bibnamefont {D\'ora}},\
  }\href {\doibase 10.1103/PhysRevB.89.161105} {\bibfield  {journal} {\bibinfo
  {journal} {Phys. Rev. B}\ }\textbf {\bibinfo {volume} {89}},\ \bibinfo
  {pages} {161105} (\bibinfo {year} {2014})}\BibitemShut {NoStop}%
\bibitem [{\citenamefont {Divakaran}\ \emph
  {et~al.}(2016{\natexlab{a}})\citenamefont {Divakaran}, \citenamefont
  {Sharma},\ and\ \citenamefont {Dutta}}]{PhysRevE.93.052133}%
  \BibitemOpen
  \bibfield  {author} {\bibinfo {author} {\bibfnamefont {U.}~\bibnamefont
  {Divakaran}}, \bibinfo {author} {\bibfnamefont {S.}~\bibnamefont {Sharma}}, \
  and\ \bibinfo {author} {\bibfnamefont {A.}~\bibnamefont {Dutta}},\ }\href
  {\doibase 10.1103/PhysRevE.93.052133} {\bibfield  {journal} {\bibinfo
  {journal} {Phys. Rev. E}\ }\textbf {\bibinfo {volume} {93}},\ \bibinfo
  {pages} {052133} (\bibinfo {year} {2016}{\natexlab{a}})}\BibitemShut
  {NoStop}%
\bibitem [{\citenamefont {Cao}\ \emph {et~al.}(2022{\natexlab{a}})\citenamefont
  {Cao}, \citenamefont {Zhong},\ and\ \citenamefont {Tong}}]{Cao202231}%
  \BibitemOpen
  \bibfield  {author} {\bibinfo {author} {\bibfnamefont {K.}~\bibnamefont
  {Cao}}, \bibinfo {author} {\bibfnamefont {M.}~\bibnamefont {Zhong}}, \ and\
  \bibinfo {author} {\bibfnamefont {P.}~\bibnamefont {Tong}},\ }\href {\doibase
  10.1088/1674-1056/ac4a6e} {\bibfield  {journal} {\bibinfo  {journal} {Chinese
  Physics B}\ }\textbf {\bibinfo {volume} {31}},\ \bibinfo {pages} {060505}
  (\bibinfo {year} {2022}{\natexlab{a}})}\BibitemShut {NoStop}%
\bibitem [{\citenamefont {Porta}\ \emph {et~al.}(2020)\citenamefont {Porta},
  \citenamefont {Cavaliere}, \citenamefont {Sassetti},\ and\ \citenamefont
  {Ziani}}]{Porta.10.1}%
  \BibitemOpen
  \bibfield  {author} {\bibinfo {author} {\bibfnamefont {S.}~\bibnamefont
  {Porta}}, \bibinfo {author} {\bibfnamefont {F.}~\bibnamefont {Cavaliere}},
  \bibinfo {author} {\bibfnamefont {M.}~\bibnamefont {Sassetti}}, \ and\
  \bibinfo {author} {\bibfnamefont {N.~T.}\ \bibnamefont {Ziani}},\ }\href
  {\doibase 10.1038/s41598-020-69621-8} {\bibfield  {journal} {\bibinfo
  {journal} {Scientific Reports}\ }\textbf {\bibinfo {volume} {10}} (\bibinfo
  {year} {2020}),\ 10.1038/s41598-020-69621-8}\BibitemShut {NoStop}%
\bibitem [{\citenamefont {Schmitt}\ and\ \citenamefont
  {Kehrein}(2015)}]{Schmitt201592}%
  \BibitemOpen
  \bibfield  {author} {\bibinfo {author} {\bibfnamefont {M.}~\bibnamefont
  {Schmitt}}\ and\ \bibinfo {author} {\bibfnamefont {S.}~\bibnamefont
  {Kehrein}},\ }\href {\doibase 10.1103/PhysRevB.92.075114} {\bibfield
  {journal} {\bibinfo  {journal} {Phys. Rev. B}\ }\textbf {\bibinfo {volume}
  {92}},\ \bibinfo {pages} {075114} (\bibinfo {year} {2015})}\BibitemShut
  {NoStop}%
\bibitem [{\citenamefont {Karrasch}\ and\ \citenamefont
  {Schuricht}(2013)}]{Karrasch201387}%
  \BibitemOpen
  \bibfield  {author} {\bibinfo {author} {\bibfnamefont {C.}~\bibnamefont
  {Karrasch}}\ and\ \bibinfo {author} {\bibfnamefont {D.}~\bibnamefont
  {Schuricht}},\ }\href {\doibase 10.1103/PhysRevB.87.195104} {\bibfield
  {journal} {\bibinfo  {journal} {Phys. Rev. B}\ }\textbf {\bibinfo {volume}
  {87}},\ \bibinfo {pages} {195104} (\bibinfo {year} {2013})}\BibitemShut
  {NoStop}%
\bibitem [{\citenamefont {Andraschko}\ and\ \citenamefont
  {Sirker}(2014)}]{Andraschko201489}%
  \BibitemOpen
  \bibfield  {author} {\bibinfo {author} {\bibfnamefont {F.}~\bibnamefont
  {Andraschko}}\ and\ \bibinfo {author} {\bibfnamefont {J.}~\bibnamefont
  {Sirker}},\ }\href {\doibase 10.1103/PhysRevB.89.125120} {\bibfield
  {journal} {\bibinfo  {journal} {Phys. Rev. B}\ }\textbf {\bibinfo {volume}
  {89}},\ \bibinfo {pages} {125120} (\bibinfo {year} {2014})}\BibitemShut
  {NoStop}%
\bibitem [{\citenamefont {Heyl}(2014)}]{Heyl2014113}%
  \BibitemOpen
  \bibfield  {author} {\bibinfo {author} {\bibfnamefont {M.}~\bibnamefont
  {Heyl}},\ }\href {\doibase 10.1103/PhysRevLett.113.205701} {\bibfield
  {journal} {\bibinfo  {journal} {Phys. Rev. Lett.}\ }\textbf {\bibinfo
  {volume} {113}},\ \bibinfo {pages} {205701} (\bibinfo {year}
  {2014})}\BibitemShut {NoStop}%
\bibitem [{\citenamefont {Kriel}\ \emph {et~al.}(2014)\citenamefont {Kriel},
  \citenamefont {Karrasch},\ and\ \citenamefont {Kehrein}}]{Kriel.90.125106}%
  \BibitemOpen
  \bibfield  {author} {\bibinfo {author} {\bibfnamefont {J.~N.}\ \bibnamefont
  {Kriel}}, \bibinfo {author} {\bibfnamefont {C.}~\bibnamefont {Karrasch}}, \
  and\ \bibinfo {author} {\bibfnamefont {S.}~\bibnamefont {Kehrein}},\ }\href
  {\doibase 10.1103/PhysRevB.90.125106} {\bibfield  {journal} {\bibinfo
  {journal} {Phys. Rev. B}\ }\textbf {\bibinfo {volume} {90}},\ \bibinfo
  {pages} {125106} (\bibinfo {year} {2014})}\BibitemShut {NoStop}%
\bibitem [{\citenamefont {Sharma}\ \emph
  {et~al.}(2015{\natexlab{a}})\citenamefont {Sharma}, \citenamefont {Suzuki},\
  and\ \citenamefont {Dutta}}]{Sharma.92.104306}%
  \BibitemOpen
  \bibfield  {author} {\bibinfo {author} {\bibfnamefont {S.}~\bibnamefont
  {Sharma}}, \bibinfo {author} {\bibfnamefont {S.}~\bibnamefont {Suzuki}}, \
  and\ \bibinfo {author} {\bibfnamefont {A.}~\bibnamefont {Dutta}},\ }\href
  {\doibase 10.1103/PhysRevB.92.104306} {\bibfield  {journal} {\bibinfo
  {journal} {Phys. Rev. B}\ }\textbf {\bibinfo {volume} {92}},\ \bibinfo
  {pages} {104306} (\bibinfo {year} {2015}{\natexlab{a}})}\BibitemShut
  {NoStop}%
\bibitem [{\citenamefont {Halimeh}\ and\ \citenamefont
  {Zauner-Stauber}(2017)}]{Halimeh201796}%
  \BibitemOpen
  \bibfield  {author} {\bibinfo {author} {\bibfnamefont {J.~C.}\ \bibnamefont
  {Halimeh}}\ and\ \bibinfo {author} {\bibfnamefont {V.}~\bibnamefont
  {Zauner-Stauber}},\ }\href {\doibase 10.1103/PhysRevB.96.134427} {\bibfield
  {journal} {\bibinfo  {journal} {Phys. Rev. B}\ }\textbf {\bibinfo {volume}
  {96}},\ \bibinfo {pages} {134427} (\bibinfo {year} {2017})}\BibitemShut
  {NoStop}%
\bibitem [{\citenamefont {Homrighausen}\ \emph {et~al.}(2017)\citenamefont
  {Homrighausen}, \citenamefont {Abeling}, \citenamefont {Zauner-Stauber},\
  and\ \citenamefont {Halimeh}}]{Homrighausen201796}%
  \BibitemOpen
  \bibfield  {author} {\bibinfo {author} {\bibfnamefont {I.}~\bibnamefont
  {Homrighausen}}, \bibinfo {author} {\bibfnamefont {N.~O.}\ \bibnamefont
  {Abeling}}, \bibinfo {author} {\bibfnamefont {V.}~\bibnamefont
  {Zauner-Stauber}}, \ and\ \bibinfo {author} {\bibfnamefont {J.~C.}\
  \bibnamefont {Halimeh}},\ }\href {\doibase 10.1103/PhysRevB.96.104436}
  {\bibfield  {journal} {\bibinfo  {journal} {Phys. Rev. B}\ }\textbf {\bibinfo
  {volume} {96}},\ \bibinfo {pages} {104436} (\bibinfo {year}
  {2017})}\BibitemShut {NoStop}%
\bibitem [{\citenamefont {Obuchi}\ \emph {et~al.}(2017)\citenamefont {Obuchi},
  \citenamefont {Suzuki},\ and\ \citenamefont
  {Takahashi}}]{PhysRevB.95.174305}%
  \BibitemOpen
  \bibfield  {author} {\bibinfo {author} {\bibfnamefont {T.}~\bibnamefont
  {Obuchi}}, \bibinfo {author} {\bibfnamefont {S.}~\bibnamefont {Suzuki}}, \
  and\ \bibinfo {author} {\bibfnamefont {K.}~\bibnamefont {Takahashi}},\ }\href
  {\doibase 10.1103/PhysRevB.95.174305} {\bibfield  {journal} {\bibinfo
  {journal} {Phys. Rev. B}\ }\textbf {\bibinfo {volume} {95}},\ \bibinfo
  {pages} {174305} (\bibinfo {year} {2017})}\BibitemShut {NoStop}%
\bibitem [{\citenamefont {Zauner-Stauber}\ and\ \citenamefont
  {Halimeh}(2017)}]{PhysRevE.96.062118}%
  \BibitemOpen
  \bibfield  {author} {\bibinfo {author} {\bibfnamefont {V.}~\bibnamefont
  {Zauner-Stauber}}\ and\ \bibinfo {author} {\bibfnamefont {J.~C.}\
  \bibnamefont {Halimeh}},\ }\href {\doibase 10.1103/PhysRevE.96.062118}
  {\bibfield  {journal} {\bibinfo  {journal} {Phys. Rev. E}\ }\textbf {\bibinfo
  {volume} {96}},\ \bibinfo {pages} {062118} (\bibinfo {year}
  {2017})}\BibitemShut {NoStop}%
\bibitem [{\citenamefont {Dutta}\ and\ \citenamefont
  {Dutta}(2017)}]{Dutta201796}%
  \BibitemOpen
  \bibfield  {author} {\bibinfo {author} {\bibfnamefont {A.}~\bibnamefont
  {Dutta}}\ and\ \bibinfo {author} {\bibfnamefont {A.}~\bibnamefont {Dutta}},\
  }\href {\doibase 10.1103/PhysRevB.96.125113} {\bibfield  {journal} {\bibinfo
  {journal} {Phys. Rev. B}\ }\textbf {\bibinfo {volume} {96}},\ \bibinfo
  {pages} {125113} (\bibinfo {year} {2017})}\BibitemShut {NoStop}%
\bibitem [{\citenamefont {\ifmmode \check{Z}\else
  \v{Z}\fi{}unkovi\ifmmode~\check{c}\else \v{c}\fi{}}\ \emph
  {et~al.}(2018)\citenamefont {\ifmmode \check{Z}\else
  \v{Z}\fi{}unkovi\ifmmode~\check{c}\else \v{c}\fi{}}, \citenamefont {Heyl},
  \citenamefont {Knap},\ and\ \citenamefont {Silva}}]{Bojan2018120}%
  \BibitemOpen
  \bibfield  {author} {\bibinfo {author} {\bibfnamefont {B.}~\bibnamefont
  {\ifmmode \check{Z}\else \v{Z}\fi{}unkovi\ifmmode~\check{c}\else
  \v{c}\fi{}}}, \bibinfo {author} {\bibfnamefont {M.}~\bibnamefont {Heyl}},
  \bibinfo {author} {\bibfnamefont {M.}~\bibnamefont {Knap}}, \ and\ \bibinfo
  {author} {\bibfnamefont {A.}~\bibnamefont {Silva}},\ }\href {\doibase
  10.1103/PhysRevLett.120.130601} {\bibfield  {journal} {\bibinfo  {journal}
  {Phys. Rev. Lett.}\ }\textbf {\bibinfo {volume} {120}},\ \bibinfo {pages}
  {130601} (\bibinfo {year} {2018})}\BibitemShut {NoStop}%
\bibitem [{\citenamefont {Halimeh}\ \emph {et~al.}(2020)\citenamefont
  {Halimeh}, \citenamefont {Van~Damme}, \citenamefont {Zauner-Stauber},\ and\
  \citenamefont {Vanderstraeten}}]{Halimeh.2.033111}%
  \BibitemOpen
  \bibfield  {author} {\bibinfo {author} {\bibfnamefont {J.~C.}\ \bibnamefont
  {Halimeh}}, \bibinfo {author} {\bibfnamefont {M.}~\bibnamefont {Van~Damme}},
  \bibinfo {author} {\bibfnamefont {V.}~\bibnamefont {Zauner-Stauber}}, \ and\
  \bibinfo {author} {\bibfnamefont {L.}~\bibnamefont {Vanderstraeten}},\ }\href
  {\doibase 10.1103/PhysRevResearch.2.033111} {\bibfield  {journal} {\bibinfo
  {journal} {Phys. Rev. Res.}\ }\textbf {\bibinfo {volume} {2}},\ \bibinfo
  {pages} {033111} (\bibinfo {year} {2020})}\BibitemShut {NoStop}%
\bibitem [{\citenamefont {Karrasch}\ and\ \citenamefont
  {Schuricht}(2017)}]{PhysRevB.95.075143}%
  \BibitemOpen
  \bibfield  {author} {\bibinfo {author} {\bibfnamefont {C.}~\bibnamefont
  {Karrasch}}\ and\ \bibinfo {author} {\bibfnamefont {D.}~\bibnamefont
  {Schuricht}},\ }\href {\doibase 10.1103/PhysRevB.95.075143} {\bibfield
  {journal} {\bibinfo  {journal} {Phys. Rev. B}\ }\textbf {\bibinfo {volume}
  {95}},\ \bibinfo {pages} {075143} (\bibinfo {year} {2017})}\BibitemShut
  {NoStop}%
\bibitem [{\citenamefont {Zhou}\ \emph {et~al.}(2018)\citenamefont {Zhou},
  \citenamefont {Wang}, \citenamefont {Wang},\ and\ \citenamefont
  {Gong}}]{Zhou201898}%
  \BibitemOpen
  \bibfield  {author} {\bibinfo {author} {\bibfnamefont {L.}~\bibnamefont
  {Zhou}}, \bibinfo {author} {\bibfnamefont {Q.-h.}\ \bibnamefont {Wang}},
  \bibinfo {author} {\bibfnamefont {H.}~\bibnamefont {Wang}}, \ and\ \bibinfo
  {author} {\bibfnamefont {J.}~\bibnamefont {Gong}},\ }\href {\doibase
  10.1103/PhysRevA.98.022129} {\bibfield  {journal} {\bibinfo  {journal} {Phys.
  Rev. A}\ }\textbf {\bibinfo {volume} {98}},\ \bibinfo {pages} {022129}
  (\bibinfo {year} {2018})}\BibitemShut {NoStop}%
\bibitem [{\citenamefont {Mondal}\ and\ \citenamefont
  {Nag}(2022)}]{Mondal.2022.106}%
  \BibitemOpen
  \bibfield  {author} {\bibinfo {author} {\bibfnamefont {D.}~\bibnamefont
  {Mondal}}\ and\ \bibinfo {author} {\bibfnamefont {T.}~\bibnamefont {Nag}},\
  }\href {\doibase 10.1103/PhysRevB.106.054308} {\bibfield  {journal} {\bibinfo
   {journal} {Phys. Rev. B}\ }\textbf {\bibinfo {volume} {106}},\ \bibinfo
  {pages} {054308} (\bibinfo {year} {2022})}\BibitemShut {NoStop}%
\bibitem [{\citenamefont {Mondal}\ and\ \citenamefont
  {Nag}(2023)}]{Mondal.2023.107}%
  \BibitemOpen
  \bibfield  {author} {\bibinfo {author} {\bibfnamefont {D.}~\bibnamefont
  {Mondal}}\ and\ \bibinfo {author} {\bibfnamefont {T.}~\bibnamefont {Nag}},\
  }\href {\doibase 10.1103/PhysRevB.107.184311} {\bibfield  {journal} {\bibinfo
   {journal} {Phys. Rev. B}\ }\textbf {\bibinfo {volume} {107}},\ \bibinfo
  {pages} {184311} (\bibinfo {year} {2023})}\BibitemShut {NoStop}%
\bibitem [{\citenamefont {Abdi}(2019)}]{Mehdi2019100}%
  \BibitemOpen
  \bibfield  {author} {\bibinfo {author} {\bibfnamefont {M.}~\bibnamefont
  {Abdi}},\ }\href {\doibase 10.1103/PhysRevB.100.184310} {\bibfield  {journal}
  {\bibinfo  {journal} {Phys. Rev. B}\ }\textbf {\bibinfo {volume} {100}},\
  \bibinfo {pages} {184310} (\bibinfo {year} {2019})}\BibitemShut {NoStop}%
\bibitem [{\citenamefont {Yang}\ \emph {et~al.}(2017)\citenamefont {Yang},
  \citenamefont {Wang}, \citenamefont {Wang}, \citenamefont {Gao},\ and\
  \citenamefont {Chen}}]{Yang201796}%
  \BibitemOpen
  \bibfield  {author} {\bibinfo {author} {\bibfnamefont {C.}~\bibnamefont
  {Yang}}, \bibinfo {author} {\bibfnamefont {Y.}~\bibnamefont {Wang}}, \bibinfo
  {author} {\bibfnamefont {P.}~\bibnamefont {Wang}}, \bibinfo {author}
  {\bibfnamefont {X.}~\bibnamefont {Gao}}, \ and\ \bibinfo {author}
  {\bibfnamefont {S.}~\bibnamefont {Chen}},\ }\href {\doibase
  10.1103/PhysRevB.95.184201} {\bibfield  {journal} {\bibinfo  {journal} {Phys.
  Rev. B}\ }\textbf {\bibinfo {volume} {95}},\ \bibinfo {pages} {184201}
  (\bibinfo {year} {2017})}\BibitemShut {NoStop}%
\bibitem [{\citenamefont {Yin}\ \emph {et~al.}(2018)\citenamefont {Yin},
  \citenamefont {Chen}, \citenamefont {Gao},\ and\ \citenamefont
  {Wang}}]{PhysRevA.97.033624}%
  \BibitemOpen
  \bibfield  {author} {\bibinfo {author} {\bibfnamefont {H.}~\bibnamefont
  {Yin}}, \bibinfo {author} {\bibfnamefont {S.}~\bibnamefont {Chen}}, \bibinfo
  {author} {\bibfnamefont {X.}~\bibnamefont {Gao}}, \ and\ \bibinfo {author}
  {\bibfnamefont {P.}~\bibnamefont {Wang}},\ }\href {\doibase
  10.1103/PhysRevA.97.033624} {\bibfield  {journal} {\bibinfo  {journal} {Phys.
  Rev. A}\ }\textbf {\bibinfo {volume} {97}},\ \bibinfo {pages} {033624}
  (\bibinfo {year} {2018})}\BibitemShut {NoStop}%
\bibitem [{\citenamefont {Mendl}\ and\ \citenamefont
  {Budich}(2019)}]{Mendl2019100}%
  \BibitemOpen
  \bibfield  {author} {\bibinfo {author} {\bibfnamefont {C.~B.}\ \bibnamefont
  {Mendl}}\ and\ \bibinfo {author} {\bibfnamefont {J.~C.}\ \bibnamefont
  {Budich}},\ }\href {\doibase 10.1103/PhysRevB.100.224307} {\bibfield
  {journal} {\bibinfo  {journal} {Phys. Rev. B}\ }\textbf {\bibinfo {volume}
  {100}},\ \bibinfo {pages} {224307} (\bibinfo {year} {2019})}\BibitemShut
  {NoStop}%
\bibitem [{\citenamefont {Cao}\ \emph {et~al.}(2020)\citenamefont {Cao},
  \citenamefont {Li}, \citenamefont {Zhong},\ and\ \citenamefont
  {Tong}}]{Cao2020102}%
  \BibitemOpen
  \bibfield  {author} {\bibinfo {author} {\bibfnamefont {K.}~\bibnamefont
  {Cao}}, \bibinfo {author} {\bibfnamefont {W.}~\bibnamefont {Li}}, \bibinfo
  {author} {\bibfnamefont {M.}~\bibnamefont {Zhong}}, \ and\ \bibinfo {author}
  {\bibfnamefont {P.}~\bibnamefont {Tong}},\ }\href {\doibase
  10.1103/PhysRevB.102.014207} {\bibfield  {journal} {\bibinfo  {journal}
  {Phys. Rev. B}\ }\textbf {\bibinfo {volume} {102}},\ \bibinfo {pages}
  {014207} (\bibinfo {year} {2020})}\BibitemShut {NoStop}%
\bibitem [{\citenamefont {Modak}\ and\ \citenamefont
  {Rakshit}(2021)}]{Modak2021103}%
  \BibitemOpen
  \bibfield  {author} {\bibinfo {author} {\bibfnamefont {R.}~\bibnamefont
  {Modak}}\ and\ \bibinfo {author} {\bibfnamefont {D.}~\bibnamefont
  {Rakshit}},\ }\href {\doibase 10.1103/PhysRevB.103.224310} {\bibfield
  {journal} {\bibinfo  {journal} {Phys. Rev. B}\ }\textbf {\bibinfo {volume}
  {103}},\ \bibinfo {pages} {224310} (\bibinfo {year} {2021})}\BibitemShut
  {NoStop}%
\bibitem [{\citenamefont {Kuliashov}\ \emph {et~al.}(2023)\citenamefont
  {Kuliashov}, \citenamefont {Markov},\ and\ \citenamefont
  {Rubtsov}}]{Kuliashov.107.094304}%
  \BibitemOpen
  \bibfield  {author} {\bibinfo {author} {\bibfnamefont {O.~N.}\ \bibnamefont
  {Kuliashov}}, \bibinfo {author} {\bibfnamefont {A.~A.}\ \bibnamefont
  {Markov}}, \ and\ \bibinfo {author} {\bibfnamefont {A.~N.}\ \bibnamefont
  {Rubtsov}},\ }\href {\doibase 10.1103/PhysRevB.107.094304} {\bibfield
  {journal} {\bibinfo  {journal} {Phys. Rev. B}\ }\textbf {\bibinfo {volume}
  {107}},\ \bibinfo {pages} {094304} (\bibinfo {year} {2023})}\BibitemShut
  {NoStop}%
\bibitem [{\citenamefont {Mishra}\ \emph {et~al.}(2020)\citenamefont {Mishra},
  \citenamefont {Jafari},\ and\ \citenamefont {Akbari}}]{Mishra.53.375301}%
  \BibitemOpen
  \bibfield  {author} {\bibinfo {author} {\bibfnamefont {U.}~\bibnamefont
  {Mishra}}, \bibinfo {author} {\bibfnamefont {R.}~\bibnamefont {Jafari}}, \
  and\ \bibinfo {author} {\bibfnamefont {A.}~\bibnamefont {Akbari}},\ }\href
  {\doibase 10.1088/1751-8121/ab97de} {\bibfield  {journal} {\bibinfo
  {journal} {Journal of Physics A: Mathematical and Theoretical}\ }\textbf
  {\bibinfo {volume} {53}},\ \bibinfo {pages} {375301} (\bibinfo {year}
  {2020})}\BibitemShut {NoStop}%
\bibitem [{\citenamefont {Yang}\ \emph {et~al.}(2019)\citenamefont {Yang},
  \citenamefont {Zhou}, \citenamefont {Ma}, \citenamefont {Kong}, \citenamefont
  {Wang}, \citenamefont {Qin}, \citenamefont {Rong}, \citenamefont {Wang},
  \citenamefont {Shi}, \citenamefont {Gong},\ and\ \citenamefont
  {Du}}]{Yang2019100}%
  \BibitemOpen
  \bibfield  {author} {\bibinfo {author} {\bibfnamefont {K.}~\bibnamefont
  {Yang}}, \bibinfo {author} {\bibfnamefont {L.}~\bibnamefont {Zhou}}, \bibinfo
  {author} {\bibfnamefont {W.}~\bibnamefont {Ma}}, \bibinfo {author}
  {\bibfnamefont {X.}~\bibnamefont {Kong}}, \bibinfo {author} {\bibfnamefont
  {P.}~\bibnamefont {Wang}}, \bibinfo {author} {\bibfnamefont {X.}~\bibnamefont
  {Qin}}, \bibinfo {author} {\bibfnamefont {X.}~\bibnamefont {Rong}}, \bibinfo
  {author} {\bibfnamefont {Y.}~\bibnamefont {Wang}}, \bibinfo {author}
  {\bibfnamefont {F.}~\bibnamefont {Shi}}, \bibinfo {author} {\bibfnamefont
  {J.}~\bibnamefont {Gong}}, \ and\ \bibinfo {author} {\bibfnamefont
  {J.}~\bibnamefont {Du}},\ }\href {\doibase 10.1103/PhysRevB.100.085308}
  {\bibfield  {journal} {\bibinfo  {journal} {Phys. Rev. B}\ }\textbf {\bibinfo
  {volume} {100}},\ \bibinfo {pages} {085308} (\bibinfo {year}
  {2019})}\BibitemShut {NoStop}%
\bibitem [{\citenamefont {Zamani}\ \emph {et~al.}(2020)\citenamefont {Zamani},
  \citenamefont {Jafari},\ and\ \citenamefont {Langari}}]{Zamani2020102}%
  \BibitemOpen
  \bibfield  {author} {\bibinfo {author} {\bibfnamefont {S.}~\bibnamefont
  {Zamani}}, \bibinfo {author} {\bibfnamefont {R.}~\bibnamefont {Jafari}}, \
  and\ \bibinfo {author} {\bibfnamefont {A.}~\bibnamefont {Langari}},\ }\href
  {\doibase 10.1103/PhysRevB.102.144306} {\bibfield  {journal} {\bibinfo
  {journal} {Phys. Rev. B}\ }\textbf {\bibinfo {volume} {102}},\ \bibinfo
  {pages} {144306} (\bibinfo {year} {2020})}\BibitemShut {NoStop}%
\bibitem [{\citenamefont {Shirai}\ \emph {et~al.}(2020)\citenamefont {Shirai},
  \citenamefont {Todo},\ and\ \citenamefont {Miyashita}}]{Shirai.101.013809}%
  \BibitemOpen
  \bibfield  {author} {\bibinfo {author} {\bibfnamefont {T.}~\bibnamefont
  {Shirai}}, \bibinfo {author} {\bibfnamefont {S.}~\bibnamefont {Todo}}, \ and\
  \bibinfo {author} {\bibfnamefont {S.}~\bibnamefont {Miyashita}},\ }\href
  {\doibase 10.1103/PhysRevA.101.013809} {\bibfield  {journal} {\bibinfo
  {journal} {Phys. Rev. A}\ }\textbf {\bibinfo {volume} {101}},\ \bibinfo
  {pages} {013809} (\bibinfo {year} {2020})}\BibitemShut {NoStop}%
\bibitem [{\citenamefont {Zhou}\ and\ \citenamefont
  {Du}(2021{\natexlab{a}})}]{Zhou202133}%
  \BibitemOpen
  \bibfield  {author} {\bibinfo {author} {\bibfnamefont {L.}~\bibnamefont
  {Zhou}}\ and\ \bibinfo {author} {\bibfnamefont {Q.}~\bibnamefont {Du}},\
  }\href {\doibase 10.1088/1361-648x/ac0b60} {\bibfield  {journal} {\bibinfo
  {journal} {Journal of Physics: Condensed Matter}\ }\textbf {\bibinfo {volume}
  {33}},\ \bibinfo {pages} {345403} (\bibinfo {year}
  {2021}{\natexlab{a}})}\BibitemShut {NoStop}%
\bibitem [{\citenamefont {Jafari}\ and\ \citenamefont
  {Akbari}(2021)}]{Jafari2021103}%
  \BibitemOpen
  \bibfield  {author} {\bibinfo {author} {\bibfnamefont {R.}~\bibnamefont
  {Jafari}}\ and\ \bibinfo {author} {\bibfnamefont {A.}~\bibnamefont
  {Akbari}},\ }\href {\doibase 10.1103/PhysRevA.103.012204} {\bibfield
  {journal} {\bibinfo  {journal} {Phys. Rev. A}\ }\textbf {\bibinfo {volume}
  {103}},\ \bibinfo {pages} {012204} (\bibinfo {year} {2021})}\BibitemShut
  {NoStop}%
\bibitem [{\citenamefont {Hamazaki}(2021)}]{NC.12.5108}%
  \BibitemOpen
  \bibfield  {author} {\bibinfo {author} {\bibfnamefont {R.}~\bibnamefont
  {Hamazaki}},\ }\href {\doibase 10.1038/s41467-021-25355-3} {\bibfield
  {journal} {\bibinfo  {journal} {Nature Communications}\ }\textbf {\bibinfo
  {volume} {12}},\ \bibinfo {pages} {5108} (\bibinfo {year}
  {2021})}\BibitemShut {NoStop}%
\bibitem [{\citenamefont {Zamani}\ \emph {et~al.}(2022)\citenamefont {Zamani},
  \citenamefont {Jafari},\ and\ \citenamefont {Langari}}]{PhysRevB.105.094304}%
  \BibitemOpen
  \bibfield  {author} {\bibinfo {author} {\bibfnamefont {S.}~\bibnamefont
  {Zamani}}, \bibinfo {author} {\bibfnamefont {R.}~\bibnamefont {Jafari}}, \
  and\ \bibinfo {author} {\bibfnamefont {A.}~\bibnamefont {Langari}},\ }\href
  {\doibase 10.1103/PhysRevB.105.094304} {\bibfield  {journal} {\bibinfo
  {journal} {Phys. Rev. B}\ }\textbf {\bibinfo {volume} {105}},\ \bibinfo
  {pages} {094304} (\bibinfo {year} {2022})}\BibitemShut {NoStop}%
\bibitem [{\citenamefont {Jafari}\ \emph {et~al.}(2022)\citenamefont {Jafari},
  \citenamefont {Akbari}, \citenamefont {Mishra},\ and\ \citenamefont
  {Johannesson}}]{Jafari2022105}%
  \BibitemOpen
  \bibfield  {author} {\bibinfo {author} {\bibfnamefont {R.}~\bibnamefont
  {Jafari}}, \bibinfo {author} {\bibfnamefont {A.}~\bibnamefont {Akbari}},
  \bibinfo {author} {\bibfnamefont {U.}~\bibnamefont {Mishra}}, \ and\ \bibinfo
  {author} {\bibfnamefont {H.}~\bibnamefont {Johannesson}},\ }\href {\doibase
  10.1103/PhysRevB.105.094311} {\bibfield  {journal} {\bibinfo  {journal}
  {Phys. Rev. B}\ }\textbf {\bibinfo {volume} {105}},\ \bibinfo {pages}
  {094311} (\bibinfo {year} {2022})}\BibitemShut {NoStop}%
\bibitem [{\citenamefont {Bhattacharya}\ \emph {et~al.}(2017)\citenamefont
  {Bhattacharya}, \citenamefont {Bandyopadhyay},\ and\ \citenamefont
  {Dutta}}]{Bhattacharya.96.180303}%
  \BibitemOpen
  \bibfield  {author} {\bibinfo {author} {\bibfnamefont {U.}~\bibnamefont
  {Bhattacharya}}, \bibinfo {author} {\bibfnamefont {S.}~\bibnamefont
  {Bandyopadhyay}}, \ and\ \bibinfo {author} {\bibfnamefont {A.}~\bibnamefont
  {Dutta}},\ }\href {\doibase 10.1103/PhysRevB.96.180303} {\bibfield  {journal}
  {\bibinfo  {journal} {Phys. Rev. B}\ }\textbf {\bibinfo {volume} {96}},\
  \bibinfo {pages} {180303} (\bibinfo {year} {2017})}\BibitemShut {NoStop}%
\bibitem [{\citenamefont {Heyl}\ and\ \citenamefont
  {Budich}(2017)}]{Heyl.96.180304}%
  \BibitemOpen
  \bibfield  {author} {\bibinfo {author} {\bibfnamefont {M.}~\bibnamefont
  {Heyl}}\ and\ \bibinfo {author} {\bibfnamefont {J.~C.}\ \bibnamefont
  {Budich}},\ }\href {\doibase 10.1103/PhysRevB.96.180304} {\bibfield
  {journal} {\bibinfo  {journal} {Phys. Rev. B}\ }\textbf {\bibinfo {volume}
  {96}},\ \bibinfo {pages} {180304} (\bibinfo {year} {2017})}\BibitemShut
  {NoStop}%
\bibitem [{\citenamefont {Lang}\ \emph
  {et~al.}(2018{\natexlab{a}})\citenamefont {Lang}, \citenamefont {Chen},
  \citenamefont {Hong},\ and\ \citenamefont {Fan}}]{Lang.98.134310}%
  \BibitemOpen
  \bibfield  {author} {\bibinfo {author} {\bibfnamefont {H.}~\bibnamefont
  {Lang}}, \bibinfo {author} {\bibfnamefont {Y.}~\bibnamefont {Chen}}, \bibinfo
  {author} {\bibfnamefont {Q.}~\bibnamefont {Hong}}, \ and\ \bibinfo {author}
  {\bibfnamefont {H.}~\bibnamefont {Fan}},\ }\href {\doibase
  10.1103/PhysRevB.98.134310} {\bibfield  {journal} {\bibinfo  {journal} {Phys.
  Rev. B}\ }\textbf {\bibinfo {volume} {98}},\ \bibinfo {pages} {134310}
  (\bibinfo {year} {2018}{\natexlab{a}})}\BibitemShut {NoStop}%
\bibitem [{\citenamefont {Bandyopadhyay}\ \emph {et~al.}(2018)\citenamefont
  {Bandyopadhyay}, \citenamefont {Laha}, \citenamefont {Bhattacharya},\ and\
  \citenamefont {Dutta}}]{Bandyopadhyay.8}%
  \BibitemOpen
  \bibfield  {author} {\bibinfo {author} {\bibfnamefont {S.}~\bibnamefont
  {Bandyopadhyay}}, \bibinfo {author} {\bibfnamefont {S.}~\bibnamefont {Laha}},
  \bibinfo {author} {\bibfnamefont {U.}~\bibnamefont {Bhattacharya}}, \ and\
  \bibinfo {author} {\bibfnamefont {A.}~\bibnamefont {Dutta}},\ }\href
  {\doibase 10.1038/s41598-018-30377-x} {\bibfield  {journal} {\bibinfo
  {journal} {Scientific Reports}\ }\textbf {\bibinfo {volume} {8}} (\bibinfo
  {year} {2018}),\ 10.1038/s41598-018-30377-x}\BibitemShut {NoStop}%
\bibitem [{\citenamefont {Hou}\ \emph {et~al.}(2020)\citenamefont {Hou},
  \citenamefont {Gao}, \citenamefont {Guo}, \citenamefont {He}, \citenamefont
  {Liu},\ and\ \citenamefont {Chien}}]{Hou.102.104305}%
  \BibitemOpen
  \bibfield  {author} {\bibinfo {author} {\bibfnamefont {X.-Y.}\ \bibnamefont
  {Hou}}, \bibinfo {author} {\bibfnamefont {Q.-C.}\ \bibnamefont {Gao}},
  \bibinfo {author} {\bibfnamefont {H.}~\bibnamefont {Guo}}, \bibinfo {author}
  {\bibfnamefont {Y.}~\bibnamefont {He}}, \bibinfo {author} {\bibfnamefont
  {T.}~\bibnamefont {Liu}}, \ and\ \bibinfo {author} {\bibfnamefont {C.-C.}\
  \bibnamefont {Chien}},\ }\href {\doibase 10.1103/PhysRevB.102.104305}
  {\bibfield  {journal} {\bibinfo  {journal} {Phys. Rev. B}\ }\textbf {\bibinfo
  {volume} {102}},\ \bibinfo {pages} {104305} (\bibinfo {year}
  {2020})}\BibitemShut {NoStop}%
\bibitem [{\citenamefont {Kyaw}\ \emph {et~al.}(2020)\citenamefont {Kyaw},
  \citenamefont {Bastidas}, \citenamefont {Tangpanitanon}, \citenamefont
  {Romero},\ and\ \citenamefont {Kwek}}]{Kyaw.101.012111}%
  \BibitemOpen
  \bibfield  {author} {\bibinfo {author} {\bibfnamefont {T.~H.}\ \bibnamefont
  {Kyaw}}, \bibinfo {author} {\bibfnamefont {V.~M.}\ \bibnamefont {Bastidas}},
  \bibinfo {author} {\bibfnamefont {J.}~\bibnamefont {Tangpanitanon}}, \bibinfo
  {author} {\bibfnamefont {G.}~\bibnamefont {Romero}}, \ and\ \bibinfo {author}
  {\bibfnamefont {L.-C.}\ \bibnamefont {Kwek}},\ }\href {\doibase
  10.1103/PhysRevA.101.012111} {\bibfield  {journal} {\bibinfo  {journal}
  {Phys. Rev. A}\ }\textbf {\bibinfo {volume} {101}},\ \bibinfo {pages}
  {012111} (\bibinfo {year} {2020})}\BibitemShut {NoStop}%
\bibitem [{\citenamefont {Mera}\ \emph {et~al.}(2018)\citenamefont {Mera},
  \citenamefont {Vlachou}, \citenamefont {Paunkovi\ifmmode~\acute{c}\else
  \'{c}\fi{}}, \citenamefont {Vieira},\ and\ \citenamefont
  {Viyuela}}]{Mera.97.094110}%
  \BibitemOpen
  \bibfield  {author} {\bibinfo {author} {\bibfnamefont {B.}~\bibnamefont
  {Mera}}, \bibinfo {author} {\bibfnamefont {C.}~\bibnamefont {Vlachou}},
  \bibinfo {author} {\bibfnamefont {N.}~\bibnamefont
  {Paunkovi\ifmmode~\acute{c}\else \'{c}\fi{}}}, \bibinfo {author}
  {\bibfnamefont {V.~R.}\ \bibnamefont {Vieira}}, \ and\ \bibinfo {author}
  {\bibfnamefont {O.}~\bibnamefont {Viyuela}},\ }\href {\doibase
  10.1103/PhysRevB.97.094110} {\bibfield  {journal} {\bibinfo  {journal} {Phys.
  Rev. B}\ }\textbf {\bibinfo {volume} {97}},\ \bibinfo {pages} {094110}
  (\bibinfo {year} {2018})}\BibitemShut {NoStop}%
\bibitem [{\citenamefont {Sedlmayr}\ \emph {et~al.}(2018)\citenamefont
  {Sedlmayr}, \citenamefont {Fleischhauer},\ and\ \citenamefont
  {Sirker}}]{Sedlmayr.97.045147}%
  \BibitemOpen
  \bibfield  {author} {\bibinfo {author} {\bibfnamefont {N.}~\bibnamefont
  {Sedlmayr}}, \bibinfo {author} {\bibfnamefont {M.}~\bibnamefont
  {Fleischhauer}}, \ and\ \bibinfo {author} {\bibfnamefont {J.}~\bibnamefont
  {Sirker}},\ }\href {\doibase 10.1103/PhysRevB.97.045147} {\bibfield
  {journal} {\bibinfo  {journal} {Phys. Rev. B}\ }\textbf {\bibinfo {volume}
  {97}},\ \bibinfo {pages} {045147} (\bibinfo {year} {2018})}\BibitemShut
  {NoStop}%
\bibitem [{\citenamefont {Link}\ and\ \citenamefont
  {Strunz}(2020)}]{Link.125.143602}%
  \BibitemOpen
  \bibfield  {author} {\bibinfo {author} {\bibfnamefont {V.}~\bibnamefont
  {Link}}\ and\ \bibinfo {author} {\bibfnamefont {W.~T.}\ \bibnamefont
  {Strunz}},\ }\href {\doibase 10.1103/PhysRevLett.125.143602} {\bibfield
  {journal} {\bibinfo  {journal} {Phys. Rev. Lett.}\ }\textbf {\bibinfo
  {volume} {125}},\ \bibinfo {pages} {143602} (\bibinfo {year}
  {2020})}\BibitemShut {NoStop}%
\bibitem [{\citenamefont {Hou}\ \emph {et~al.}(2021)\citenamefont {Hou},
  \citenamefont {Guo},\ and\ \citenamefont {Chien}}]{Hou.104.023303}%
  \BibitemOpen
  \bibfield  {author} {\bibinfo {author} {\bibfnamefont {X.-Y.}\ \bibnamefont
  {Hou}}, \bibinfo {author} {\bibfnamefont {H.}~\bibnamefont {Guo}}, \ and\
  \bibinfo {author} {\bibfnamefont {C.-C.}\ \bibnamefont {Chien}},\ }\href
  {\doibase 10.1103/PhysRevA.104.023303} {\bibfield  {journal} {\bibinfo
  {journal} {Phys. Rev. A}\ }\textbf {\bibinfo {volume} {104}},\ \bibinfo
  {pages} {023303} (\bibinfo {year} {2021})}\BibitemShut {NoStop}%
\bibitem [{\citenamefont {Heyl}(2015)}]{Heyl.115.140602}%
  \BibitemOpen
  \bibfield  {author} {\bibinfo {author} {\bibfnamefont {M.}~\bibnamefont
  {Heyl}},\ }\href {\doibase 10.1103/PhysRevLett.115.140602} {\bibfield
  {journal} {\bibinfo  {journal} {Phys. Rev. Lett.}\ }\textbf {\bibinfo
  {volume} {115}},\ \bibinfo {pages} {140602} (\bibinfo {year}
  {2015})}\BibitemShut {NoStop}%
\bibitem [{\citenamefont {Vajna}\ and\ \citenamefont
  {D\'ora}(2015)}]{Vajna.91.155127}%
  \BibitemOpen
  \bibfield  {author} {\bibinfo {author} {\bibfnamefont {S.}~\bibnamefont
  {Vajna}}\ and\ \bibinfo {author} {\bibfnamefont {B.}~\bibnamefont {D\'ora}},\
  }\href {\doibase 10.1103/PhysRevB.91.155127} {\bibfield  {journal} {\bibinfo
  {journal} {Phys. Rev. B}\ }\textbf {\bibinfo {volume} {91}},\ \bibinfo
  {pages} {155127} (\bibinfo {year} {2015})}\BibitemShut {NoStop}%
\bibitem [{\citenamefont {Puskarov}\ and\ \citenamefont
  {Schuricht}(2016)}]{Tatjana.1.1.003}%
  \BibitemOpen
  \bibfield  {author} {\bibinfo {author} {\bibfnamefont {T.}~\bibnamefont
  {Puskarov}}\ and\ \bibinfo {author} {\bibfnamefont {D.}~\bibnamefont
  {Schuricht}},\ }\href {\doibase 10.21468/SciPostPhys.1.1.003} {\bibfield
  {journal} {\bibinfo  {journal} {SciPost Phys.}\ }\textbf {\bibinfo {volume}
  {1}},\ \bibinfo {pages} {003} (\bibinfo {year} {2016})}\BibitemShut {NoStop}%
\bibitem [{\citenamefont {Lang}\ \emph
  {et~al.}(2018{\natexlab{b}})\citenamefont {Lang}, \citenamefont {Frank},\
  and\ \citenamefont {Halimeh}}]{Lang.121.130603}%
  \BibitemOpen
  \bibfield  {author} {\bibinfo {author} {\bibfnamefont {J.}~\bibnamefont
  {Lang}}, \bibinfo {author} {\bibfnamefont {B.}~\bibnamefont {Frank}}, \ and\
  \bibinfo {author} {\bibfnamefont {J.~C.}\ \bibnamefont {Halimeh}},\ }\href
  {\doibase 10.1103/PhysRevLett.121.130603} {\bibfield  {journal} {\bibinfo
  {journal} {Phys. Rev. Lett.}\ }\textbf {\bibinfo {volume} {121}},\ \bibinfo
  {pages} {130603} (\bibinfo {year} {2018}{\natexlab{b}})}\BibitemShut
  {NoStop}%
\bibitem [{\citenamefont {Huang}\ \emph {et~al.}(2019)\citenamefont {Huang},
  \citenamefont {Banerjee},\ and\ \citenamefont {Heyl}}]{Huang.122.250401}%
  \BibitemOpen
  \bibfield  {author} {\bibinfo {author} {\bibfnamefont {Y.-P.}\ \bibnamefont
  {Huang}}, \bibinfo {author} {\bibfnamefont {D.}~\bibnamefont {Banerjee}}, \
  and\ \bibinfo {author} {\bibfnamefont {M.}~\bibnamefont {Heyl}},\ }\href
  {\doibase 10.1103/PhysRevLett.122.250401} {\bibfield  {journal} {\bibinfo
  {journal} {Phys. Rev. Lett.}\ }\textbf {\bibinfo {volume} {122}},\ \bibinfo
  {pages} {250401} (\bibinfo {year} {2019})}\BibitemShut {NoStop}%
\bibitem [{\citenamefont {Jafari}\ \emph {et~al.}(2019)\citenamefont {Jafari},
  \citenamefont {Johannesson}, \citenamefont {Langari},\ and\ \citenamefont
  {Martin-Delgado}}]{Jafari.99.054302}%
  \BibitemOpen
  \bibfield  {author} {\bibinfo {author} {\bibfnamefont {R.}~\bibnamefont
  {Jafari}}, \bibinfo {author} {\bibfnamefont {H.}~\bibnamefont {Johannesson}},
  \bibinfo {author} {\bibfnamefont {A.}~\bibnamefont {Langari}}, \ and\
  \bibinfo {author} {\bibfnamefont {M.~A.}\ \bibnamefont {Martin-Delgado}},\
  }\href {\doibase 10.1103/PhysRevB.99.054302} {\bibfield  {journal} {\bibinfo
  {journal} {Phys. Rev. B}\ }\textbf {\bibinfo {volume} {99}},\ \bibinfo
  {pages} {054302} (\bibinfo {year} {2019})}\BibitemShut {NoStop}%
\bibitem [{\citenamefont {Khatun}\ and\ \citenamefont
  {Bhattacharjee}(2019)}]{Khatun.123.160603}%
  \BibitemOpen
  \bibfield  {author} {\bibinfo {author} {\bibfnamefont {A.}~\bibnamefont
  {Khatun}}\ and\ \bibinfo {author} {\bibfnamefont {S.~M.}\ \bibnamefont
  {Bhattacharjee}},\ }\href {\doibase 10.1103/PhysRevLett.123.160603}
  {\bibfield  {journal} {\bibinfo  {journal} {Phys. Rev. Lett.}\ }\textbf
  {\bibinfo {volume} {123}},\ \bibinfo {pages} {160603} (\bibinfo {year}
  {2019})}\BibitemShut {NoStop}%
\bibitem [{\citenamefont {Lahiri}\ and\ \citenamefont
  {Bera}(2019)}]{Lahiri.99.174311}%
  \BibitemOpen
  \bibfield  {author} {\bibinfo {author} {\bibfnamefont {A.}~\bibnamefont
  {Lahiri}}\ and\ \bibinfo {author} {\bibfnamefont {S.}~\bibnamefont {Bera}},\
  }\href {\doibase 10.1103/PhysRevB.99.174311} {\bibfield  {journal} {\bibinfo
  {journal} {Phys. Rev. B}\ }\textbf {\bibinfo {volume} {99}},\ \bibinfo
  {pages} {174311} (\bibinfo {year} {2019})}\BibitemShut {NoStop}%
\bibitem [{\citenamefont {Liu}\ and\ \citenamefont
  {Guo}(2019)}]{Liu.99.104307}%
  \BibitemOpen
  \bibfield  {author} {\bibinfo {author} {\bibfnamefont {T.}~\bibnamefont
  {Liu}}\ and\ \bibinfo {author} {\bibfnamefont {H.}~\bibnamefont {Guo}},\
  }\href {\doibase 10.1103/PhysRevB.99.104307} {\bibfield  {journal} {\bibinfo
  {journal} {Phys. Rev. B}\ }\textbf {\bibinfo {volume} {99}},\ \bibinfo
  {pages} {104307} (\bibinfo {year} {2019})}\BibitemShut {NoStop}%
\bibitem [{\citenamefont {Srivastav}\ \emph {et~al.}(2019)\citenamefont
  {Srivastav}, \citenamefont {Bhattacharya},\ and\ \citenamefont
  {Dutta}}]{Srivastav.100.144203}%
  \BibitemOpen
  \bibfield  {author} {\bibinfo {author} {\bibfnamefont {V.}~\bibnamefont
  {Srivastav}}, \bibinfo {author} {\bibfnamefont {U.}~\bibnamefont
  {Bhattacharya}}, \ and\ \bibinfo {author} {\bibfnamefont {A.}~\bibnamefont
  {Dutta}},\ }\href {\doibase 10.1103/PhysRevB.100.144203} {\bibfield
  {journal} {\bibinfo  {journal} {Phys. Rev. B}\ }\textbf {\bibinfo {volume}
  {100}},\ \bibinfo {pages} {144203} (\bibinfo {year} {2019})}\BibitemShut
  {NoStop}%
\bibitem [{\citenamefont {Gul\'acsi}\ \emph {et~al.}(2020)\citenamefont
  {Gul\'acsi}, \citenamefont {Heyl},\ and\ \citenamefont
  {D\'ora}}]{Gulacsi.101.205135}%
  \BibitemOpen
  \bibfield  {author} {\bibinfo {author} {\bibfnamefont {B.}~\bibnamefont
  {Gul\'acsi}}, \bibinfo {author} {\bibfnamefont {M.}~\bibnamefont {Heyl}}, \
  and\ \bibinfo {author} {\bibfnamefont {B.}~\bibnamefont {D\'ora}},\ }\href
  {\doibase 10.1103/PhysRevB.101.205135} {\bibfield  {journal} {\bibinfo
  {journal} {Phys. Rev. B}\ }\textbf {\bibinfo {volume} {101}},\ \bibinfo
  {pages} {205135} (\bibinfo {year} {2020})}\BibitemShut {NoStop}%
\bibitem [{\citenamefont {Meibohm}\ and\ \citenamefont
  {Esposito}(2023)}]{Meibohm_2023.023034}%
  \BibitemOpen
  \bibfield  {author} {\bibinfo {author} {\bibfnamefont {J.}~\bibnamefont
  {Meibohm}}\ and\ \bibinfo {author} {\bibfnamefont {M.}~\bibnamefont
  {Esposito}},\ }\href {\doibase 10.1088/1367-2630/acbc41} {\bibfield
  {journal} {\bibinfo  {journal} {New Journal of Physics}\ }\textbf {\bibinfo
  {volume} {25}},\ \bibinfo {pages} {023034} (\bibinfo {year}
  {2023})}\BibitemShut {NoStop}%
\bibitem [{\citenamefont {Wong}\ and\ \citenamefont
  {Yu}(2022)}]{Wong.105.174307}%
  \BibitemOpen
  \bibfield  {author} {\bibinfo {author} {\bibfnamefont {C.~Y.}\ \bibnamefont
  {Wong}}\ and\ \bibinfo {author} {\bibfnamefont {W.~C.}\ \bibnamefont {Yu}},\
  }\href {\doibase 10.1103/PhysRevB.105.174307} {\bibfield  {journal} {\bibinfo
   {journal} {Phys. Rev. B}\ }\textbf {\bibinfo {volume} {105}},\ \bibinfo
  {pages} {174307} (\bibinfo {year} {2022})}\BibitemShut {NoStop}%
\bibitem [{\citenamefont {Wrze\ifmmode~\acute{s}\else \'{s}\fi{}niewski}\ \emph
  {et~al.}(2022)\citenamefont {Wrze\ifmmode~\acute{s}\else \'{s}\fi{}niewski},
  \citenamefont {Weymann}, \citenamefont {Sedlmayr},\ and\ \citenamefont
  {Doma\ifmmode~\acute{n}\else \'{n}\fi{}ski}}]{Wrzessniewski.105.094514}%
  \BibitemOpen
  \bibfield  {author} {\bibinfo {author} {\bibfnamefont {K.}~\bibnamefont
  {Wrze\ifmmode~\acute{s}\else \'{s}\fi{}niewski}}, \bibinfo {author}
  {\bibfnamefont {I.}~\bibnamefont {Weymann}}, \bibinfo {author} {\bibfnamefont
  {N.}~\bibnamefont {Sedlmayr}}, \ and\ \bibinfo {author} {\bibfnamefont
  {T.}~\bibnamefont {Doma\ifmmode~\acute{n}\else \'{n}\fi{}ski}},\ }\href
  {\doibase 10.1103/PhysRevB.105.094514} {\bibfield  {journal} {\bibinfo
  {journal} {Phys. Rev. B}\ }\textbf {\bibinfo {volume} {105}},\ \bibinfo
  {pages} {094514} (\bibinfo {year} {2022})}\BibitemShut {NoStop}%
\bibitem [{\citenamefont {Hashizume}\ \emph {et~al.}(2022)\citenamefont
  {Hashizume}, \citenamefont {McCulloch},\ and\ \citenamefont
  {Halimeh}}]{Hashizume.4.013250}%
  \BibitemOpen
  \bibfield  {author} {\bibinfo {author} {\bibfnamefont {T.}~\bibnamefont
  {Hashizume}}, \bibinfo {author} {\bibfnamefont {I.~P.}\ \bibnamefont
  {McCulloch}}, \ and\ \bibinfo {author} {\bibfnamefont {J.~C.}\ \bibnamefont
  {Halimeh}},\ }\href {\doibase 10.1103/PhysRevResearch.4.013250} {\bibfield
  {journal} {\bibinfo  {journal} {Phys. Rev. Res.}\ }\textbf {\bibinfo {volume}
  {4}},\ \bibinfo {pages} {013250} (\bibinfo {year} {2022})}\BibitemShut
  {NoStop}%
\bibitem [{\citenamefont {Fl\"{a}schner}\ \emph {et~al.}(2017)\citenamefont
  {Fl\"{a}schner}, \citenamefont {Vogel}, \citenamefont {Tarnowski},
  \citenamefont {Rem}, \citenamefont {L\"{u}hmann}, \citenamefont {Heyl},
  \citenamefont {Budich}, \citenamefont {Mathey}, \citenamefont {Sengstock},\
  and\ \citenamefont {Weitenberg}}]{Vogel201714}%
  \BibitemOpen
  \bibfield  {author} {\bibinfo {author} {\bibfnamefont {N.}~\bibnamefont
  {Fl\"{a}schner}}, \bibinfo {author} {\bibfnamefont {D.}~\bibnamefont
  {Vogel}}, \bibinfo {author} {\bibfnamefont {M.}~\bibnamefont {Tarnowski}},
  \bibinfo {author} {\bibfnamefont {B.~S.}\ \bibnamefont {Rem}}, \bibinfo
  {author} {\bibfnamefont {D.~S.}\ \bibnamefont {L\"{u}hmann}}, \bibinfo
  {author} {\bibfnamefont {M.}~\bibnamefont {Heyl}}, \bibinfo {author}
  {\bibfnamefont {J.~C.}\ \bibnamefont {Budich}}, \bibinfo {author}
  {\bibfnamefont {L.}~\bibnamefont {Mathey}}, \bibinfo {author} {\bibfnamefont
  {K.}~\bibnamefont {Sengstock}}, \ and\ \bibinfo {author} {\bibfnamefont
  {C.}~\bibnamefont {Weitenberg}},\ }\href {\doibase 10.1038/s41567-017-0013-8}
  {\bibfield  {journal} {\bibinfo  {journal} {Nature Physics}\ }\textbf
  {\bibinfo {volume} {14}},\ \bibinfo {pages} {265} (\bibinfo {year}
  {2017})}\BibitemShut {NoStop}%
\bibitem [{\citenamefont {Jurcevic}\ \emph {et~al.}(2017)\citenamefont
  {Jurcevic}, \citenamefont {Shen}, \citenamefont {Hauke}, \citenamefont
  {Maier}, \citenamefont {Brydges}, \citenamefont {Hempel}, \citenamefont
  {Lanyon}, \citenamefont {Heyl}, \citenamefont {Blatt},\ and\ \citenamefont
  {Roos}}]{Jurcevic2017119}%
  \BibitemOpen
  \bibfield  {author} {\bibinfo {author} {\bibfnamefont {P.}~\bibnamefont
  {Jurcevic}}, \bibinfo {author} {\bibfnamefont {H.}~\bibnamefont {Shen}},
  \bibinfo {author} {\bibfnamefont {P.}~\bibnamefont {Hauke}}, \bibinfo
  {author} {\bibfnamefont {C.}~\bibnamefont {Maier}}, \bibinfo {author}
  {\bibfnamefont {T.}~\bibnamefont {Brydges}}, \bibinfo {author} {\bibfnamefont
  {C.}~\bibnamefont {Hempel}}, \bibinfo {author} {\bibfnamefont {B.~P.}\
  \bibnamefont {Lanyon}}, \bibinfo {author} {\bibfnamefont {M.}~\bibnamefont
  {Heyl}}, \bibinfo {author} {\bibfnamefont {R.}~\bibnamefont {Blatt}}, \ and\
  \bibinfo {author} {\bibfnamefont {C.~F.}\ \bibnamefont {Roos}},\ }\href
  {\doibase 10.1103/PhysRevLett.119.080501} {\bibfield  {journal} {\bibinfo
  {journal} {Phys. Rev. Lett.}\ }\textbf {\bibinfo {volume} {119}},\ \bibinfo
  {pages} {080501} (\bibinfo {year} {2017})}\BibitemShut {NoStop}%
\bibitem [{\citenamefont {Chen}\ \emph {et~al.}(2020)\citenamefont {Chen},
  \citenamefont {Cui}, \citenamefont {Ai}, \citenamefont {He}, \citenamefont
  {Huang}, \citenamefont {Han}, \citenamefont {Li},\ and\ \citenamefont
  {Guo}}]{Chen.102.042222}%
  \BibitemOpen
  \bibfield  {author} {\bibinfo {author} {\bibfnamefont {Z.}~\bibnamefont
  {Chen}}, \bibinfo {author} {\bibfnamefont {J.-M.}\ \bibnamefont {Cui}},
  \bibinfo {author} {\bibfnamefont {M.-Z.}\ \bibnamefont {Ai}}, \bibinfo
  {author} {\bibfnamefont {R.}~\bibnamefont {He}}, \bibinfo {author}
  {\bibfnamefont {Y.-F.}\ \bibnamefont {Huang}}, \bibinfo {author}
  {\bibfnamefont {Y.-J.}\ \bibnamefont {Han}}, \bibinfo {author} {\bibfnamefont
  {C.-F.}\ \bibnamefont {Li}}, \ and\ \bibinfo {author} {\bibfnamefont {G.-C.}\
  \bibnamefont {Guo}},\ }\href {\doibase 10.1103/PhysRevA.102.042222}
  {\bibfield  {journal} {\bibinfo  {journal} {Phys. Rev. A}\ }\textbf {\bibinfo
  {volume} {102}},\ \bibinfo {pages} {042222} (\bibinfo {year}
  {2020})}\BibitemShut {NoStop}%
\bibitem [{\citenamefont {Muniz}\ \emph {et~al.}(2020)\citenamefont {Muniz},
  \citenamefont {Barberena}, \citenamefont {Lewis-Swan}, \citenamefont {Young},
  \citenamefont {Cline}, \citenamefont {Rey},\ and\ \citenamefont
  {Thompson}}]{Muniz.580.602}%
  \BibitemOpen
  \bibfield  {author} {\bibinfo {author} {\bibfnamefont {J.~A.}\ \bibnamefont
  {Muniz}}, \bibinfo {author} {\bibfnamefont {D.}~\bibnamefont {Barberena}},
  \bibinfo {author} {\bibfnamefont {R.~J.}\ \bibnamefont {Lewis-Swan}},
  \bibinfo {author} {\bibfnamefont {D.~J.}\ \bibnamefont {Young}}, \bibinfo
  {author} {\bibfnamefont {J.~R.~K.}\ \bibnamefont {Cline}}, \bibinfo {author}
  {\bibfnamefont {A.~M.}\ \bibnamefont {Rey}}, \ and\ \bibinfo {author}
  {\bibfnamefont {J.~K.}\ \bibnamefont {Thompson}},\ }\href {\doibase
  10.1038/s41586-020-2224-x} {\bibfield  {journal} {\bibinfo  {journal}
  {Nature}\ }\textbf {\bibinfo {volume} {580}},\ \bibinfo {pages} {602}
  (\bibinfo {year} {2020})}\BibitemShut {NoStop}%
\bibitem [{\citenamefont {Zhang}\ \emph {et~al.}(2017)\citenamefont {Zhang},
  \citenamefont {Pagano}, \citenamefont {Hess}, \citenamefont {Kyprianidis},
  \citenamefont {Ecker}, \citenamefont {Kaplan}, \citenamefont {Gorshkov},
  \citenamefont {Gong},\ and\ \citenamefont {Monroe}}]{Zhang2017551}%
  \BibitemOpen
  \bibfield  {author} {\bibinfo {author} {\bibfnamefont {J.}~\bibnamefont
  {Zhang}}, \bibinfo {author} {\bibfnamefont {G.}~\bibnamefont {Pagano}},
  \bibinfo {author} {\bibfnamefont {P.~W.}\ \bibnamefont {Hess}}, \bibinfo
  {author} {\bibfnamefont {A.}~\bibnamefont {Kyprianidis}}, \bibinfo {author}
  {\bibfnamefont {P.~B.}\ \bibnamefont {Ecker}}, \bibinfo {author}
  {\bibfnamefont {H.}~\bibnamefont {Kaplan}}, \bibinfo {author} {\bibfnamefont
  {A.~V.}\ \bibnamefont {Gorshkov}}, \bibinfo {author} {\bibfnamefont {Z.~X.}\
  \bibnamefont {Gong}}, \ and\ \bibinfo {author} {\bibfnamefont
  {C.}~\bibnamefont {Monroe}},\ }\href {\doibase 10.1038/nature24654}
  {\bibfield  {journal} {\bibinfo  {journal} {Nature}\ }\textbf {\bibinfo
  {volume} {551}},\ \bibinfo {pages} {601} (\bibinfo {year}
  {2017})}\BibitemShut {NoStop}%
\bibitem [{\citenamefont {Nie}\ \emph {et~al.}(2020)\citenamefont {Nie},
  \citenamefont {Wei}, \citenamefont {Chen}, \citenamefont {Zhang},
  \citenamefont {Zhao}, \citenamefont {Qiu}, \citenamefont {Tian},
  \citenamefont {Ji}, \citenamefont {Xin}, \citenamefont {Lu},\ and\
  \citenamefont {Li}}]{Nie2020124}%
  \BibitemOpen
  \bibfield  {author} {\bibinfo {author} {\bibfnamefont {X.}~\bibnamefont
  {Nie}}, \bibinfo {author} {\bibfnamefont {B.-B.}\ \bibnamefont {Wei}},
  \bibinfo {author} {\bibfnamefont {X.}~\bibnamefont {Chen}}, \bibinfo {author}
  {\bibfnamefont {Z.}~\bibnamefont {Zhang}}, \bibinfo {author} {\bibfnamefont
  {X.}~\bibnamefont {Zhao}}, \bibinfo {author} {\bibfnamefont {C.}~\bibnamefont
  {Qiu}}, \bibinfo {author} {\bibfnamefont {Y.}~\bibnamefont {Tian}}, \bibinfo
  {author} {\bibfnamefont {Y.}~\bibnamefont {Ji}}, \bibinfo {author}
  {\bibfnamefont {T.}~\bibnamefont {Xin}}, \bibinfo {author} {\bibfnamefont
  {D.}~\bibnamefont {Lu}}, \ and\ \bibinfo {author} {\bibfnamefont
  {J.}~\bibnamefont {Li}},\ }\href {\doibase 10.1103/PhysRevLett.124.250601}
  {\bibfield  {journal} {\bibinfo  {journal} {Phys. Rev. Lett.}\ }\textbf
  {\bibinfo {volume} {124}},\ \bibinfo {pages} {250601} (\bibinfo {year}
  {2020})}\BibitemShut {NoStop}%
\bibitem [{\citenamefont {Wang}\ \emph {et~al.}(2019)\citenamefont {Wang},
  \citenamefont {Qiu}, \citenamefont {Xiao}, \citenamefont {Zhan},
  \citenamefont {Bian}, \citenamefont {Yi},\ and\ \citenamefont
  {Xue}}]{Wang2019122}%
  \BibitemOpen
  \bibfield  {author} {\bibinfo {author} {\bibfnamefont {K.}~\bibnamefont
  {Wang}}, \bibinfo {author} {\bibfnamefont {X.}~\bibnamefont {Qiu}}, \bibinfo
  {author} {\bibfnamefont {L.}~\bibnamefont {Xiao}}, \bibinfo {author}
  {\bibfnamefont {X.}~\bibnamefont {Zhan}}, \bibinfo {author} {\bibfnamefont
  {Z.}~\bibnamefont {Bian}}, \bibinfo {author} {\bibfnamefont {W.}~\bibnamefont
  {Yi}}, \ and\ \bibinfo {author} {\bibfnamefont {P.}~\bibnamefont {Xue}},\
  }\href {\doibase 10.1103/PhysRevLett.122.020501} {\bibfield  {journal}
  {\bibinfo  {journal} {Phys. Rev. Lett.}\ }\textbf {\bibinfo {volume} {122}},\
  \bibinfo {pages} {020501} (\bibinfo {year} {2019})}\BibitemShut {NoStop}%
\bibitem [{\citenamefont {Xu}\ \emph {et~al.}(2020)\citenamefont {Xu},
  \citenamefont {Wang}, \citenamefont {Heyl}, \citenamefont {Budich},
  \citenamefont {Pan}, \citenamefont {Chen}, \citenamefont {Jan}, \citenamefont
  {Sun}, \citenamefont {Xu}, \citenamefont {Han}, \citenamefont {Li},\ and\
  \citenamefont {Guo}}]{Xu20209}%
  \BibitemOpen
  \bibfield  {author} {\bibinfo {author} {\bibfnamefont {X.~Y.}\ \bibnamefont
  {Xu}}, \bibinfo {author} {\bibfnamefont {Q.~Q.}\ \bibnamefont {Wang}},
  \bibinfo {author} {\bibfnamefont {M.}~\bibnamefont {Heyl}}, \bibinfo {author}
  {\bibfnamefont {J.~C.}\ \bibnamefont {Budich}}, \bibinfo {author}
  {\bibfnamefont {W.~W.}\ \bibnamefont {Pan}}, \bibinfo {author} {\bibfnamefont
  {Z.}~\bibnamefont {Chen}}, \bibinfo {author} {\bibfnamefont {M.}~\bibnamefont
  {Jan}}, \bibinfo {author} {\bibfnamefont {K.}~\bibnamefont {Sun}}, \bibinfo
  {author} {\bibfnamefont {J.~S.}\ \bibnamefont {Xu}}, \bibinfo {author}
  {\bibfnamefont {Y.~J.}\ \bibnamefont {Han}}, \bibinfo {author} {\bibfnamefont
  {C.~F.}\ \bibnamefont {Li}}, \ and\ \bibinfo {author} {\bibfnamefont {G.~C.}\
  \bibnamefont {Guo}},\ }\href {\doibase 10.1038/s41377-019-0237-8} {\bibfield
  {journal} {\bibinfo  {journal} {Light-Science Applications}\ }\textbf
  {\bibinfo {volume} {9}} (\bibinfo {year} {2020}),\
  10.1038/s41377-019-0237-8}\BibitemShut {NoStop}%
\bibitem [{\citenamefont {Tian}\ \emph {et~al.}(2020)\citenamefont {Tian},
  \citenamefont {Yang}, \citenamefont {Qiu}, \citenamefont {Liang},
  \citenamefont {Yang}, \citenamefont {Xu},\ and\ \citenamefont
  {Duan}}]{Tian2020124}%
  \BibitemOpen
  \bibfield  {author} {\bibinfo {author} {\bibfnamefont {T.}~\bibnamefont
  {Tian}}, \bibinfo {author} {\bibfnamefont {H.-X.}\ \bibnamefont {Yang}},
  \bibinfo {author} {\bibfnamefont {L.-Y.}\ \bibnamefont {Qiu}}, \bibinfo
  {author} {\bibfnamefont {H.-Y.}\ \bibnamefont {Liang}}, \bibinfo {author}
  {\bibfnamefont {Y.-B.}\ \bibnamefont {Yang}}, \bibinfo {author}
  {\bibfnamefont {Y.}~\bibnamefont {Xu}}, \ and\ \bibinfo {author}
  {\bibfnamefont {L.-M.}\ \bibnamefont {Duan}},\ }\href {\doibase
  10.1103/PhysRevLett.124.043001} {\bibfield  {journal} {\bibinfo  {journal}
  {Phys. Rev. Lett.}\ }\textbf {\bibinfo {volume} {124}},\ \bibinfo {pages}
  {043001} (\bibinfo {year} {2020})}\BibitemShut {NoStop}%
\bibitem [{\citenamefont {Yuzbashyan}\ \emph {et~al.}(2006)\citenamefont
  {Yuzbashyan}, \citenamefont {Tsyplyatyev},\ and\ \citenamefont
  {Altshuler}}]{Yuzbashyan200696}%
  \BibitemOpen
  \bibfield  {author} {\bibinfo {author} {\bibfnamefont {E.~A.}\ \bibnamefont
  {Yuzbashyan}}, \bibinfo {author} {\bibfnamefont {O.}~\bibnamefont
  {Tsyplyatyev}}, \ and\ \bibinfo {author} {\bibfnamefont {B.~L.}\ \bibnamefont
  {Altshuler}},\ }\href {\doibase 10.1103/PhysRevLett.96.097005} {\bibfield
  {journal} {\bibinfo  {journal} {Phys. Rev. Lett.}\ }\textbf {\bibinfo
  {volume} {96}},\ \bibinfo {pages} {097005} (\bibinfo {year}
  {2006})}\BibitemShut {NoStop}%
\bibitem [{\citenamefont {Barmettler}\ \emph {et~al.}(2009)\citenamefont
  {Barmettler}, \citenamefont {Punk}, \citenamefont {Gritsev}, \citenamefont
  {Demler},\ and\ \citenamefont {Altman}}]{Barmettler2009102}%
  \BibitemOpen
  \bibfield  {author} {\bibinfo {author} {\bibfnamefont {P.}~\bibnamefont
  {Barmettler}}, \bibinfo {author} {\bibfnamefont {M.}~\bibnamefont {Punk}},
  \bibinfo {author} {\bibfnamefont {V.}~\bibnamefont {Gritsev}}, \bibinfo
  {author} {\bibfnamefont {E.}~\bibnamefont {Demler}}, \ and\ \bibinfo {author}
  {\bibfnamefont {E.}~\bibnamefont {Altman}},\ }\href {\doibase
  10.1103/PhysRevLett.102.130603} {\bibfield  {journal} {\bibinfo  {journal}
  {Phys. Rev. Lett.}\ }\textbf {\bibinfo {volume} {102}},\ \bibinfo {pages}
  {130603} (\bibinfo {year} {2009})}\BibitemShut {NoStop}%
\bibitem [{\citenamefont {Eckstein}\ \emph {et~al.}(2009)\citenamefont
  {Eckstein}, \citenamefont {Kollar},\ and\ \citenamefont
  {Werner}}]{Eckstein2009103}%
  \BibitemOpen
  \bibfield  {author} {\bibinfo {author} {\bibfnamefont {M.}~\bibnamefont
  {Eckstein}}, \bibinfo {author} {\bibfnamefont {M.}~\bibnamefont {Kollar}}, \
  and\ \bibinfo {author} {\bibfnamefont {P.}~\bibnamefont {Werner}},\ }\href
  {\doibase 10.1103/PhysRevLett.103.056403} {\bibfield  {journal} {\bibinfo
  {journal} {Phys. Rev. Lett.}\ }\textbf {\bibinfo {volume} {103}},\ \bibinfo
  {pages} {056403} (\bibinfo {year} {2009})}\BibitemShut {NoStop}%
\bibitem [{\citenamefont {Sciolla}\ and\ \citenamefont
  {Biroli}(2010)}]{Sciolla2010105}%
  \BibitemOpen
  \bibfield  {author} {\bibinfo {author} {\bibfnamefont {B.}~\bibnamefont
  {Sciolla}}\ and\ \bibinfo {author} {\bibfnamefont {G.}~\bibnamefont
  {Biroli}},\ }\href {\doibase 10.1103/PhysRevLett.105.220401} {\bibfield
  {journal} {\bibinfo  {journal} {Phys. Rev. Lett.}\ }\textbf {\bibinfo
  {volume} {105}},\ \bibinfo {pages} {220401} (\bibinfo {year}
  {2010})}\BibitemShut {NoStop}%
\bibitem [{\citenamefont {Dziarmaga}(2010)}]{Dziarmaga201059}%
  \BibitemOpen
  \bibfield  {author} {\bibinfo {author} {\bibfnamefont {J.}~\bibnamefont
  {Dziarmaga}},\ }\href {\doibase 10.1080/00018732.2010.514702} {\bibfield
  {journal} {\bibinfo  {journal} {Advances in Physics}\ }\textbf {\bibinfo
  {volume} {59}},\ \bibinfo {pages} {1063} (\bibinfo {year}
  {2010})}\BibitemShut {NoStop}%
\bibitem [{\citenamefont {Budich}\ and\ \citenamefont
  {Heyl}(2016)}]{Budich201693}%
  \BibitemOpen
  \bibfield  {author} {\bibinfo {author} {\bibfnamefont {J.~C.}\ \bibnamefont
  {Budich}}\ and\ \bibinfo {author} {\bibfnamefont {M.}~\bibnamefont {Heyl}},\
  }\href {\doibase 10.1103/PhysRevB.93.085416} {\bibfield  {journal} {\bibinfo
  {journal} {Phys. Rev. B}\ }\textbf {\bibinfo {volume} {93}},\ \bibinfo
  {pages} {085416} (\bibinfo {year} {2016})}\BibitemShut {NoStop}%
\bibitem [{\citenamefont {Huang}\ and\ \citenamefont
  {Balatsky}(2016)}]{Huang2016117}%
  \BibitemOpen
  \bibfield  {author} {\bibinfo {author} {\bibfnamefont {Z.}~\bibnamefont
  {Huang}}\ and\ \bibinfo {author} {\bibfnamefont {A.~V.}\ \bibnamefont
  {Balatsky}},\ }\href {\doibase 10.1103/PhysRevLett.117.086802} {\bibfield
  {journal} {\bibinfo  {journal} {Phys. Rev. Lett.}\ }\textbf {\bibinfo
  {volume} {117}},\ \bibinfo {pages} {086802} (\bibinfo {year}
  {2016})}\BibitemShut {NoStop}%
\bibitem [{\citenamefont {Haldar}\ \emph {et~al.}(2020)\citenamefont {Haldar},
  \citenamefont {Roy}, \citenamefont {Chanda}, \citenamefont {Sen(De)},\ and\
  \citenamefont {Sen}}]{Haldar2020101}%
  \BibitemOpen
  \bibfield  {author} {\bibinfo {author} {\bibfnamefont {S.}~\bibnamefont
  {Haldar}}, \bibinfo {author} {\bibfnamefont {S.}~\bibnamefont {Roy}},
  \bibinfo {author} {\bibfnamefont {T.}~\bibnamefont {Chanda}}, \bibinfo
  {author} {\bibfnamefont {A.}~\bibnamefont {Sen(De)}}, \ and\ \bibinfo
  {author} {\bibfnamefont {U.}~\bibnamefont {Sen}},\ }\href {\doibase
  10.1103/PhysRevB.101.224304} {\bibfield  {journal} {\bibinfo  {journal}
  {Phys. Rev. B}\ }\textbf {\bibinfo {volume} {101}},\ \bibinfo {pages}
  {224304} (\bibinfo {year} {2020})}\BibitemShut {NoStop}%
\bibitem [{\citenamefont {Mas\l{}owski}\ and\ \citenamefont
  {Sedlmayr}(2020)}]{Sedlmayr2020}%
  \BibitemOpen
  \bibfield  {author} {\bibinfo {author} {\bibfnamefont {T.}~\bibnamefont
  {Mas\l{}owski}}\ and\ \bibinfo {author} {\bibfnamefont {N.}~\bibnamefont
  {Sedlmayr}},\ }\href {\doibase 10.1103/PhysRevB.101.014301} {\bibfield
  {journal} {\bibinfo  {journal} {Phys. Rev. B}\ }\textbf {\bibinfo {volume}
  {101}},\ \bibinfo {pages} {014301} (\bibinfo {year} {2020})}\BibitemShut
  {NoStop}%
\bibitem [{\citenamefont {Altland}\ and\ \citenamefont
  {Zirnbauer}(1997)}]{Altland199755}%
  \BibitemOpen
  \bibfield  {author} {\bibinfo {author} {\bibfnamefont {A.}~\bibnamefont
  {Altland}}\ and\ \bibinfo {author} {\bibfnamefont {M.~R.}\ \bibnamefont
  {Zirnbauer}},\ }\href {\doibase 10.1103/PhysRevB.55.1142} {\bibfield
  {journal} {\bibinfo  {journal} {Phys. Rev. B}\ }\textbf {\bibinfo {volume}
  {55}},\ \bibinfo {pages} {1142} (\bibinfo {year} {1997})}\BibitemShut
  {NoStop}%
\bibitem [{\citenamefont {Pfeuty}(1979)}]{Pfeuty1979245}%
  \BibitemOpen
  \bibfield  {author} {\bibinfo {author} {\bibfnamefont {P.}~\bibnamefont
  {Pfeuty}},\ }\href {\doibase https://doi.org/10.1016/0375-9601(79)90017-3}
  {\bibfield  {journal} {\bibinfo  {journal} {Phys. Lett. A}\ }\textbf
  {\bibinfo {volume} {72}},\ \bibinfo {pages} {245} (\bibinfo {year}
  {1979})}\BibitemShut {NoStop}%
\bibitem [{\citenamefont {Kitaev}(2001)}]{Kitaev_2001}%
  \BibitemOpen
  \bibfield  {author} {\bibinfo {author} {\bibfnamefont {A.~Y.}\ \bibnamefont
  {Kitaev}},\ }\href {\doibase 10.1070/1063-7869/44/10S/S29} {\bibfield
  {journal} {\bibinfo  {journal} {Physics-Uspekhi}\ }\textbf {\bibinfo {volume}
  {44}},\ \bibinfo {pages} {131} (\bibinfo {year} {2001})}\BibitemShut
  {NoStop}%
\bibitem [{\citenamefont {Sharma}\ \emph
  {et~al.}(2015{\natexlab{b}})\citenamefont {Sharma}, \citenamefont {Suzuki},\
  and\ \citenamefont {Dutta}}]{sharma2015}%
  \BibitemOpen
  \bibfield  {author} {\bibinfo {author} {\bibfnamefont {S.}~\bibnamefont
  {Sharma}}, \bibinfo {author} {\bibfnamefont {S.}~\bibnamefont {Suzuki}}, \
  and\ \bibinfo {author} {\bibfnamefont {A.}~\bibnamefont {Dutta}},\ }\href
  {\doibase 10.1103/PhysRevB.92.104306} {\bibfield  {journal} {\bibinfo
  {journal} {Phys. Rev. B}\ }\textbf {\bibinfo {volume} {92}},\ \bibinfo
  {pages} {104306} (\bibinfo {year} {2015}{\natexlab{b}})}\BibitemShut
  {NoStop}%
\bibitem [{\citenamefont {Sharma}\ \emph {et~al.}(2016)\citenamefont {Sharma},
  \citenamefont {Divakaran}, \citenamefont {Polkovnikov},\ and\ \citenamefont
  {Dutta}}]{Sharma201693}%
  \BibitemOpen
  \bibfield  {author} {\bibinfo {author} {\bibfnamefont {S.}~\bibnamefont
  {Sharma}}, \bibinfo {author} {\bibfnamefont {U.}~\bibnamefont {Divakaran}},
  \bibinfo {author} {\bibfnamefont {A.}~\bibnamefont {Polkovnikov}}, \ and\
  \bibinfo {author} {\bibfnamefont {A.}~\bibnamefont {Dutta}},\ }\href
  {\doibase 10.1103/PhysRevB.93.144306} {\bibfield  {journal} {\bibinfo
  {journal} {Phys. Rev. B}\ }\textbf {\bibinfo {volume} {93}},\ \bibinfo
  {pages} {144306} (\bibinfo {year} {2016})}\BibitemShut {NoStop}%
\bibitem [{\citenamefont {Zhang}\ and\ \citenamefont
  {Yang}(2016)}]{Zhang_2016114}%
  \BibitemOpen
  \bibfield  {author} {\bibinfo {author} {\bibfnamefont {J.~M.}\ \bibnamefont
  {Zhang}}\ and\ \bibinfo {author} {\bibfnamefont {H.-T.}\ \bibnamefont
  {Yang}},\ }\href {\doibase 10.1209/0295-5075/114/60001} {\bibfield  {journal}
  {\bibinfo  {journal} {Europhysics Letters}\ }\textbf {\bibinfo {volume}
  {114}},\ \bibinfo {pages} {60001} (\bibinfo {year} {2016})}\BibitemShut
  {NoStop}%
\bibitem [{\citenamefont {Divakaran}\ \emph
  {et~al.}(2016{\natexlab{b}})\citenamefont {Divakaran}, \citenamefont
  {Sharma},\ and\ \citenamefont {Dutta}}]{Divakaran201693}%
  \BibitemOpen
  \bibfield  {author} {\bibinfo {author} {\bibfnamefont {U.}~\bibnamefont
  {Divakaran}}, \bibinfo {author} {\bibfnamefont {S.}~\bibnamefont {Sharma}}, \
  and\ \bibinfo {author} {\bibfnamefont {A.}~\bibnamefont {Dutta}},\ }\href
  {\doibase 10.1103/PhysRevE.93.052133} {\bibfield  {journal} {\bibinfo
  {journal} {Phys. Rev. E}\ }\textbf {\bibinfo {volume} {93}},\ \bibinfo
  {pages} {052133} (\bibinfo {year} {2016}{\natexlab{b}})}\BibitemShut
  {NoStop}%
\bibitem [{Note1()}]{Note1}%
  \BibitemOpen
  \bibinfo {note} {It is easy to check that at the critical wave vector $k_{c}$
  and the first critical time $t_{0}^{*}$, the initial state is given by
  $\protect \mathrm {|\psi _{0}\delimiter "526930B
  =p_{k_{c}1}|u_{k_{c}1}^{f}\delimiter "526930B
  +p_{k_{c}2}|u_{k_{c}2}^{f}\delimiter "526930B }$, and the time-evolved state
  is $\protect \mathrm {|\psi (t)\delimiter "526930B
  =i(p_{k_{c}1}|u_{k_{c}1}^{f}\delimiter "526930B
  -p_{k_{c}2}|u_{k_{c}2}^{f}\delimiter "526930B )}$. Therefore, it is necessary
  that $\protect \mathrm {\delimiter "426830A \psi _{0}|\psi (t)\delimiter
  "526930B =i(|p_{k_{c}1}|^{2}-|p_{k_{c}2}|^{2})=0}$ in order to satisfy the
  condition of the DQPT.}\BibitemShut {Stop}%
\bibitem [{\citenamefont {Berry}(1984)}]{Berry1984392}%
  \BibitemOpen
  \bibfield  {author} {\bibinfo {author} {\bibfnamefont {M.~V.}\ \bibnamefont
  {Berry}},\ }\href {\doibase 10.1098/rspa.1984.0023} {\bibfield  {journal}
  {\bibinfo  {journal} {Proc. R. Soc. Lond. A}\ }\textbf {\bibinfo {volume}
  {392}},\ \bibinfo {pages} {45} (\bibinfo {year} {1984})}\BibitemShut
  {NoStop}%
\bibitem [{\citenamefont {Samuel}\ and\ \citenamefont
  {Bhandari}(1988)}]{Samuel198860}%
  \BibitemOpen
  \bibfield  {author} {\bibinfo {author} {\bibfnamefont {J.}~\bibnamefont
  {Samuel}}\ and\ \bibinfo {author} {\bibfnamefont {R.}~\bibnamefont
  {Bhandari}},\ }\href {\doibase 10.1103/PhysRevLett.60.2339} {\bibfield
  {journal} {\bibinfo  {journal} {Phys. Rev. Lett.}\ }\textbf {\bibinfo
  {volume} {60}},\ \bibinfo {pages} {2339} (\bibinfo {year}
  {1988})}\BibitemShut {NoStop}%
\bibitem [{\citenamefont {Lang}\ \emph
  {et~al.}(2018{\natexlab{c}})\citenamefont {Lang}, \citenamefont {Chen},
  \citenamefont {Hong},\ and\ \citenamefont {Fan}}]{Lang201898}%
  \BibitemOpen
  \bibfield  {author} {\bibinfo {author} {\bibfnamefont {H.}~\bibnamefont
  {Lang}}, \bibinfo {author} {\bibfnamefont {Y.}~\bibnamefont {Chen}}, \bibinfo
  {author} {\bibfnamefont {Q.}~\bibnamefont {Hong}}, \ and\ \bibinfo {author}
  {\bibfnamefont {H.}~\bibnamefont {Fan}},\ }\href {\doibase
  10.1103/PhysRevB.98.134310} {\bibfield  {journal} {\bibinfo  {journal} {Phys.
  Rev. B}\ }\textbf {\bibinfo {volume} {98}},\ \bibinfo {pages} {134310}
  (\bibinfo {year} {2018}{\natexlab{c}})}\BibitemShut {NoStop}%
\bibitem [{\citenamefont {Qiu}\ \emph {et~al.}(2018)\citenamefont {Qiu},
  \citenamefont {Deng}, \citenamefont {Guo},\ and\ \citenamefont
  {Yi}}]{Qiu201898}%
  \BibitemOpen
  \bibfield  {author} {\bibinfo {author} {\bibfnamefont {X.}~\bibnamefont
  {Qiu}}, \bibinfo {author} {\bibfnamefont {T.-S.}\ \bibnamefont {Deng}},
  \bibinfo {author} {\bibfnamefont {G.-C.}\ \bibnamefont {Guo}}, \ and\
  \bibinfo {author} {\bibfnamefont {W.}~\bibnamefont {Yi}},\ }\href {\doibase
  10.1103/PhysRevA.98.021601} {\bibfield  {journal} {\bibinfo  {journal} {Phys.
  Rev. A}\ }\textbf {\bibinfo {volume} {98}},\ \bibinfo {pages} {021601}
  (\bibinfo {year} {2018})}\BibitemShut {NoStop}%
\bibitem [{\citenamefont {Ding}(2020)}]{Ding2020102}%
  \BibitemOpen
  \bibfield  {author} {\bibinfo {author} {\bibfnamefont {C.}~\bibnamefont
  {Ding}},\ }\href {\doibase 10.1103/PhysRevB.102.060409} {\bibfield  {journal}
  {\bibinfo  {journal} {Phys. Rev. B}\ }\textbf {\bibinfo {volume} {102}},\
  \bibinfo {pages} {060409} (\bibinfo {year} {2020})}\BibitemShut {NoStop}%
\bibitem [{\citenamefont {Cao}\ \emph {et~al.}()\citenamefont {Cao},
  \citenamefont {Yang}, \citenamefont {Hu},\ and\ \citenamefont
  {Yang}}]{Cao2023}%
  \BibitemOpen
  \bibfield  {author} {\bibinfo {author} {\bibfnamefont {K.}~\bibnamefont
  {Cao}}, \bibinfo {author} {\bibfnamefont {S.}~\bibnamefont {Yang}}, \bibinfo
  {author} {\bibfnamefont {Y.}~\bibnamefont {Hu}}, \ and\ \bibinfo {author}
  {\bibfnamefont {G.}~\bibnamefont {Yang}},\ }\href {\doibase
  10.48550/arXiv.2211.15976} {\bibfield  {journal} {\bibinfo  {journal} {arXiv:
  2211.15976}\ }10.48550/arXiv.2211.15976}\BibitemShut {NoStop}%
\bibitem [{\citenamefont {Zhou}\ and\ \citenamefont
  {Du}(2021{\natexlab{b}})}]{Zhou202123}%
  \BibitemOpen
  \bibfield  {author} {\bibinfo {author} {\bibfnamefont {L.}~\bibnamefont
  {Zhou}}\ and\ \bibinfo {author} {\bibfnamefont {Q.}~\bibnamefont {Du}},\
  }\href {\doibase 10.1088/1367-2630/ac0574} {\bibfield  {journal} {\bibinfo
  {journal} {New Journal of Physics}\ }\textbf {\bibinfo {volume} {23}},\
  \bibinfo {pages} {063041} (\bibinfo {year} {2021}{\natexlab{b}})}\BibitemShut
  {NoStop}%
\bibitem [{\citenamefont {Cao}\ \emph {et~al.}(2022{\natexlab{b}})\citenamefont
  {Cao}, \citenamefont {Zhong},\ and\ \citenamefont {Tong}}]{Cao202236}%
  \BibitemOpen
  \bibfield  {author} {\bibinfo {author} {\bibfnamefont {K.}~\bibnamefont
  {Cao}}, \bibinfo {author} {\bibfnamefont {M.}~\bibnamefont {Zhong}}, \ and\
  \bibinfo {author} {\bibfnamefont {P.}~\bibnamefont {Tong}},\ }\href {\doibase
  10.1088/1751-8121/ac8324} {\bibfield  {journal} {\bibinfo  {journal} {J.
  Phys. A-Math. Theor.}\ }\textbf {\bibinfo {volume} {55}},\ \bibinfo {pages}
  {365001} (\bibinfo {year} {2022}{\natexlab{b}})}\BibitemShut {NoStop}%
\end{thebibliography}%

\end{document}